\documentclass[fleqn,usenatbib]{mnras}

\usepackage{subfig}
\usepackage{comment}
\usepackage{longtable}
\usepackage{caption}
\usepackage{threeparttable}
\usepackage{booktabs}
\usepackage[figuresright]{rotating}
\newcommand{\be}{\begin{equation}}
\newcommand{\ee}{\end{equation}}
\newcommand{\bea}{\begin{eqnarray}}
\newcommand{\eea}{\end{eqnarray}}
\newcommand{\hunit}{$\rm{km \ s^{-1} \ Mpc^{-1}}$}
\newcommand{\lcdm}{$\Lambda$CDM}
\newcommand{\pcdm}{$\phi$CDM}
\usepackage{array}
\makeatletter
\newcommand{\thickhline}{%
    \noalign {\ifnum 0=`}\fi \hrule height 1pt
    \futurelet \reserved@a \@xhline
}
\newcolumntype{"}{@{\hskip\tabcolsep\vrule width 1pt\hskip\tabcolsep}}
\makeatother

\newcommand{\hiig}{H\,\textsc{ii}G}
\newcommand{\hii}{H\,\textsc{ii}}
\newcommand{\Om}{\Omega_{\rm m_0}}
\newcommand{\Ok}{\Omega_{\rm k_0}}
\newcommand{\om}{$\Omega_{\rm m_0}$}
\newcommand{\ok}{$\Omega_{\rm k_0}$}
\newcommand{\wx}{$w_{\rm X}$}
\newcommand{\obh}{\Omega_{\rm b_0}\!h^2}
\newcommand{\och}{\Omega_{\rm c_0}\!h^2}
\newcommand{\obhs}{$\Omega_{\rm b_0}\!h^2$}
\newcommand{\ochs}{$\Omega_{\rm c_0}\!h^2$}

\usepackage[T1]{fontenc}

\DeclareRobustCommand{\VAN}[3]{#2}
\let\VANthebibliography\thebibliography
\def\thebibliography{\DeclareRobustCommand{\VAN}[3]{##3}\VANthebibliography}

\usepackage{graphicx}	% Including figure files
\usepackage{amsmath}	% Advanced maths commands
\usepackage{amssymb}	% Extra maths symbols
\usepackage{fnpct}
\title[H\,\textsc{ii} galaxy, quasar, and other constraints]{Cosmological constraints from H\,\textsc{ii} starburst galaxy, quasar angular size, and other measurements}

\author[S. Cao, J. Ryan and B. Ratra]{
Shulei Cao,$^{1}$\thanks{E-mail: shulei@phys.ksu.edu}
Joseph Ryan,$^{2}$\thanks{E-mail: jwryan@mail.smu.edu}
Bharat Ratra$^{1}$\thanks{E-mail: ratra@phys.ksu.edu}
\\
$^{1}$Department of Physics, Kansas State University, 116 Cardwell Hall, Manhattan, KS 66502, USA\\
$^{2}$Department of Physics, Southern Methodist University, Dallas, TX 75275, USA
}

\date{Accepted XXX. Received YYY; in original form ZZZ}

\pubyear{2021}

\begin{document}
\label{firstpage}
\pagerange{\pageref{firstpage}--\pageref{lastpage}}
\maketitle

% Abstract of the paper
\begin{abstract}
We compare the constraints from two (2019 and 2021) compilations of H\,\textsc{ii} starburst galaxy (H\,\textsc{ii}G) data and test the model-independence of quasar angular size (QSO) data using six spatially flat and non-flat cosmological models. We find that the new 2021 compilation of H\,\textsc{ii}G data generally provides tighter constraints and prefers lower values of cosmological parameters than those from the 2019 H\,\textsc{ii}G data. QSO data by themselves give relatively model-independent constraints on the characteristic linear size, $l_{\rm m}$, of the QSOs within the sample. We also use Hubble parameter ($H(z)$), baryon acoustic oscillation (BAO), Pantheon Type Ia supernova (SN Ia) apparent magnitude (SN-Pantheon), and DES-3yr binned SN Ia apparent magnitude (SN-DES) measurements to perform joint analyses with H\,\textsc{ii}G and QSO angular size data, since their constraints are not mutually inconsistent within the six cosmological models we study. A joint analysis of $H(z)$, BAO, SN-Pantheon, SN-DES, QSO, and the newest compilation of H\,\textsc{ii}G data provides almost model-independent summary estimates of the Hubble constant, $H_0=69.7\pm1.2\ \rm{km \ s^{-1} \ Mpc^{-1}}$, the non-relativistic matter density parameter, $\Omega_{\rm m_0}=0.293\pm0.021$, and $l_{\rm m}=10.93\pm0.25$ pc.
\end{abstract}

% Select between one and six entries from the list of approved keywords.
% Don't make up new ones.

%%%%%%%%%%%%%%%%%%%%%%%%%%%%%%%%%%%%%%%%%%%%%%%%%%

%%%%%%%%%%%%%%%%% BODY OF PAPER %%%%%%%%%%%%%%%%%%

\section{Introduction} \label{sec:intro}

Many observations indicate that the Universe is currently in a phase of accelerated expansion, however, the theory behind this is not yet well-established. Although the spatially flat \lcdm\ model\footnote{The flat \lcdm\ model has flat spatial hypersurfaces and a time-independent dark energy, a cosmological constant $\Lambda$, that provides approximately 70\% of the current cosmological energy budget. Non-relativistic cold dark matter (CDM) accounts for approximately 25\% and non-relativistic baryonic matter accounts for almost all of the remaining $\sim 5\%$ of the energy budget.} \citep{peeb84} is consistent with most observations (see e.g.\ \citealp{Farooq_Ranjeet_Crandall_Ratra_2017,scolnic_et_al_2018,planck2018b,eBOSS_2020}), some potential observational discrepancies and theoretical puzzles (see e.g.\ \citealp{DiValentinoetal2021b, PerivolaropoulosSkara2021}) suggest that there still is room for other cosmological models, including, for example, non-flat \lcdm\ (the \citealp{planck2018b} cosmic microwave background (CMB) anisotropy TT,TE,EE+lowE+lensing data favor positive spatial curvature) as well as dynamical dark energy. These discrepancies and puzzles motivate us to also study dynamical dark energy models and spatially non-flat models in this paper.

In this paper we use the new \cite{GM2021} \hii\ starburst galaxy (\hiig) measured fluxes and inferred absolute luminosities (from their correlation with their measured ionized gas velocity dispersions) as standard candles to constrain cosmological models.\footnote{Our analyses of the earlier \cite{G-M_2019} data are described in \cite{Caoetal_2020} (also see \citealp{Caoetal_2021a,Caoetal_2021b,Johnsonetal2021}). For analyses of earlier \hiig\ data, see  \cite{Melnick_2000,Siegel_2005,Plionis_2011,Mania_2012,Chavez_2014,Terlevich_2015,Chavez_2016}, and references therein. For recent analyses of the \cite{GM2021} \hiig\ measurements, see \cite{Tsiapietal2021, Mehrabietal2021}.} These new \hiig\ data reach to a slightly higher redshift $z \sim 2.5$, somewhat higher than the baryon acoustic oscillation (BAO) standard ruler data that reach to $z \sim 2.3$, that we also use in this paper. In order to determine the expansion rate and geometry of the Universe, it is vital to measure distances using either standard candles or standard rules. More data sets probing wider redshift regions would provide more information and make more contributions to a better understanding of our Universe, so it is worthwhile to seek additional standard rulers. The angular sizes of quasars (QSOs) provide one such additional probe, reaching to $z \sim 2.7$, which we have explored in previous work \citep{Ryanetal2019, Caoetal_2020, Caoetal_2021a, Caoetal_2021b}. As described in those papers and in Sec. \ref{sec:data} below, intermediate luminosity QSOs have, over a fairly wide range of redshifts ($0.46 \lesssim z \lesssim 2.7$), very similar intrinsic lengths $l_{\rm m} = 11.03 \pm 0.25$ pc \citep{Cao_et_al2017b}. A knowledge of this intrinsic length scale, combined with measurements of the angular sizes of these QSOs allows one to determine the angular diameter distance out to the redshifts of the QSOs.

The QSO data that we have used in the past (from \citealp{Cao_et_al2017b}) have the following drawback, however: $l_{\rm m}$ was determined with a Gaussian process interpolation \citep{Seikel_Clarkson_Smith_2012} of the Hubble parameter from Hubble parameter ($H(z)$) data (as described in \citealp{Cao_et_al2017b}), many of which we have used in our previous analyses and also use in this paper. We have discussed this correlation between our QSO data and our $H(z)$ data in the past \citep{Ryan_powerlaw, Caoetal_2021b}, although we made the assumption in our earlier analyses that the correlation is not significant enough to have a strong effect on our results (owing to the weakness of the constraints from QSO data). Here we sidestep this problem by treating $l_{\rm m}$ as a free (nuisance) parameter, thereby constraining its value directly from our analysis. As discussed in Sec. \ref{sec:results}, we find that the value of $l_{\rm m}$ is almost independent of cosmological model, and is consistent with the value $l_{\rm m} = 11.03 \pm 0.25$ pc from \cite{Cao_et_al2017b} that we used in our earlier work. This finding suggests that these QSOs are close to being standard rulers, and it validates the result of \cite{Cao_et_al2017b}, independently of their method.

Significant constraints on cosmology now largely come from only a few data sets, at low $z \lesssim 2.3$, including BAO, Type Ia supernova (SN Ia), and $H(z)$ measurements, and at $z \sim 1100$ from CMB anisotropy observations. As mentioned above, it is useful and important to develop new probes, especially in the intermediate $ 2.3 \lesssim z \lesssim 1100$ redshift range. \hiig\ is an example, as are QSO angular sizes that have been under discussion for a longer time (see e.g.\  \citealp{gurvits_kellermann_frey_1999, vishwakarma_2001, lima_alcaniz_2002, zhu_fujimoto_2002, Chen_Ratra_2003}) with the compilation of \cite{Cao_et_al2017b} being a significant step forward. Other probes under development now include reverberation-measured Mg\,\textsc{ii} time-lag radius-luminosity relation QSOs that reach to $z \sim 1.9$ (\citealp{MartinezAldama2019, Czernyetal2021, Zajaceketal2021, Yuetal2021, Khadkaetal_2021a}). High redshift options include QSO X-ray and UV flux measurements which extend to $z \sim 7.5$ (\citealp{RisalitiandLusso_2015, RisalitiandLusso_2019, Khadka_2020a, Khadka_2020b, KhadkaRatra2021a, KhadkaRatra2021b, Yang_Banerjee_Colgain_2020, Lusso_etal_2020, Li_etal_2021, Lian_etal_2021}),\footnote{However the current QSO compilation is standardizable up to only $z \sim 1.5$--1.7 (\citealp{KhadkaRatra2021b, KhadkaRatra2021a}).} and gamma-ray burst (GRB) data that extend to $z \sim 8.2$ (\citealp{Amati2008, Amati2019, samushia_ratra_2010, Wang_2016, Demianskietal_2021, Dirirsa2019, Khadka_2020c, Khadkaetal_2021b, Wangetal2021, Huetal2021}).\footnote{Only a smaller sample of 118 GRBs is reliable enough to be used for cosmological purposes, but include GRBs that probe to $z \sim 8.2$ (\citealp{Khadka_2020c, Khadkaetal_2021b}).} As of now, all five of these probes provide mostly only weak cosmological constraints, but new data should yield tighter constraints that have the potential to soon usefully probe the largely unexplored $2 \lesssim z \lesssim 8$ part of cosmological redshift space.

Our comparisons here between the constraints from the new \cite{GM2021} data and the old \cite{G-M_2019} data show that the new data provide more restrictive constraints on most cosmological parameters. As noted above, QSO angular size data provide relatively cosmological model-independent estimates of $l_{\rm m}$. We find that the cosmological constraints from $H(z)$, BAO, SN Ia, QSO, and the new \hiig\ measurements are not mutually inconsistent, thus we combine them to provide more restrictive constraints on the cosmological and nuisance parameters. The almost model-independent summary constraints from this data combination are measurements of the Hubble constant, $H_0=69.7\pm1.2$ \hunit, the non-relativistic matter density parameter, $\Omega_{\rm m_0}=0.293\pm0.021$, and the QSO characteristic linear size, $l_{\rm m}=10.93\pm0.25$ pc. The estimate of $H_0$ is in better agreement with the median statistics estimate of \cite{chenratmed} ($H_0 = 68 \pm 2.8$ \hunit) than with the measurements of \cite{planck2018b} ($H_0 = 67.4 \pm 0.5$ \hunit) and \cite{Riess_2021} ($H_0 = 73.2 \pm 1.3$ \hunit). Although the most-favored model is the spatially-flat \lcdm\ model, there is room for some mild dark energy dynamics and a little non-zero spatial curvature energy density. We also find that currently accelerating cosmological expansion is favored by most of the data combinations we study (except for QSO data alone).

This paper is organized as follows. The models we study are briefly described in Section \ref{sec:model}. The data we used are introduced in Section \ref{sec:data} with the data analysis method presented in Section \ref{sec:analysis}. We summarize our results and conclusions in Sections \ref{sec:results} and \ref{sec:conclusion}.

\section{Cosmological models}
\label{sec:model}

We use various combinations of observational data to constrain the cosmological parameters of six spatially-flat and non-flat \lcdm, XCDM, and \pcdm\ models and study the goodness of fit.\footnote{For recent observational constraints on spatial curvature see \citet{Farooqetal2015}, \citet{Chenetal2016}, \citet{Ranaetal2017}, \citet{Oobaetal2018a, Oobaetal2018b}, \citet{Yuetal2018}, \citet{ParkRatra2019a, ParkRatra2019b}, \citet{Wei2018}, \citet{DESCollaboration2019}, \citet{Lietal2020}, \citet{Handley2019}, \citet{EfstathiouGratton2020}, \citet{DiValentinoetal2021a}, \citet{VelasquezToribioFabris2020}, \citet{Vagnozzietal2020, Vagnozzietal2021}, \citet{KiDSCollaboration2021}, \citet{ArjonaNesseris2021}, \citet{Dhawanetal2021}, and references therein.} The main features of the models we use are summarized below. We assume a minimal neutrino sector, with three massless neutrino species, with the effective number of relativistic neutrino species $N_{\rm eff} = 3.046$. We neglect the late-time contribution of non-relativistic neutrinos and treat the baryonic (\obhs) and cold dark matter (\ochs) energy density parameters as free cosmological parameters to be determined from the data. The non-relativistic matter density parameter $\Om = (\obh + \och)/{h^2}$ is a derived parameter.

In the \lcdm\ models, the expansion rate function $E(z) \equiv H(z)/H_0$ as a function of redshift $z$ is
\begin{equation}
\label{eq:E(z)_LCDM}
    E(z) = \sqrt{\Om\left(1 + z\right)^3 + \Ok\left(1 + z\right)^2 + \Omega_{\Lambda}},
\end{equation}
where 
\begin{equation}
    \Omega_{\Lambda} = 1 - \Om - \Ok,
\end{equation}
with $\Ok$ being the curvature energy density parameter. There are four free parameters: $h$, \obhs\!, \ochs\!, and \ok\ in the non-flat \lcdm\ case and three in the flat case where $\Ok = 0$.

In the XCDM parametrizations, the expansion rate function is
\begin{equation}
    \label{eq:E(z)_NFXCDM}
    E(z) = \sqrt{\Om\left(1 + z\right)^3 + \Ok\left(1 + z\right)^2 + \Omega_{\rm X_0}\left(1 + z\right)^{3\left(1 + w_{\rm X}\right)}},
\end{equation}
where $w_{\rm X}$ is the equation of state parameter of the X-fluid, and
\begin{equation}
    \Omega_{\rm X_0} = 1 - \Om - \Ok.
\end{equation}
There are five free parameters: $h$, \obhs\!, \ochs\!, \ok\!, and $w_{\rm X}$ in the non-flat XCDM case and four in the flat case where $\Ok = 0$.

In the \pcdm\ models \citep{peebrat88,ratpeeb88,pavlov13}\footnote{For recent observational constraints on the $\phi$CDM model see \citet{Avsajanishvilietal2015}, \citet{SolaPeracaulaetal2018, SolaPercaulaetal2019}, \citet{Zhaietal2017}, \citet{Oobaetal2018c, Oobaetal2019}, \citet{ParkRatra2018, ParkRatra2019c, ParkRatra2020}, \citet{Sangwanetal2018}, \citet{Singhetal2019}, \citet{UrenaLopezRoy2020}, \citet{SinhaBanerjee2021}, and references therein.}, the expansion rate function is
\begin{equation}
    \label{eq:E(z)_NFphiCDM}
    E(z) = \sqrt{\Om\left(1 + z\right)^3 + \Ok\left(1 + z\right)^2 + \Omega_{\phi}(z,\alpha)},
\end{equation}
where the energy density parameter of the scalar field $\phi$, $\Omega_{\phi}(z,\alpha)$, is determined by simultaneously numerically integrating the scalar field's equation of motion
\be\label{em}
\ddot{\phi}+3\bigg(\frac{\dot{a}}{a}\bigg)\dot{\phi}+V'(\phi)=0,
\ee 
with a potential energy density
\be\label{PE}
V(\phi)=\frac{1}{2}\kappa m_p^2\phi^{-\alpha},
\ee
and the Friedmann equation (\ref{eq:E(z)_NFphiCDM}) where $H_0 E(z) = \dot{a}/a$. In these equations $a$ is the scale factor and an overdot denotes a time derivative, the prime denotes a derivative with respect to the argument, $m_p$ is the Planck mass, the parameter $\alpha \geq 0$, and the constant $\kappa$ can be treated as a shooting parameter which is determined by the shooting method implemented in the Cosmic Linear Anisotropy Solving System (\textsc{class}) code \citep{class}. There are five free parameters: $h$, \obhs\!, \ochs\!, \ok\!, and $\alpha$ in the non-flat \pcdm\ case and four in the flat case where $\Ok = 0$.

\section{Data}
\label{sec:data}

In this paper our main focus is on a new set of \hiig\ data (\citealp{GM2021}, which we dub ``\hiig-2021''). We compare cosmological constraints from these \hiig-2021 data to those from earlier \hiig\ data. We also use these \hiig-2021 data and BAO, $H(z)$, SN Ia, and QSO angular size measurements to constrain cosmological parameters in the models we study.

The 31 $H(z)$ measurements we use, that span the redshift range $0.070 \leq z \leq 1.965$, are given in Table 2 of \cite{Ryan_1}.\footnote{These measurements were taken from \cite{69}, \cite{71}, \cite{70}, \cite{73}, \cite{72}, \cite{moresco_et_al_2016}, and \cite{15}.} The 11 BAO measurements we use, that span the redshift range $0.38 \leq z \leq 2.334$, are listed in Table 1 of \cite{Caoetal_2021b}.\footnote{These measurements were taken from \cite{Alam_et_al_2017}, \cite{3}, \cite{Carter_2018}, \cite{DES_2019b}, and \cite{duMas2020}.} Information on systematic errors of these data can be found in \cite{Caoetal_2021a}.

The SN-Pantheon data we use consist of 1048 SN Ia measurements, spanning the redshift range $0.01<z<2.3$, compiled in \cite{scolnic_et_al_2018}. The SN-DES data we use consist of 20 binned measurements (of 207 SN Ia measurements), spanning the redshift range $0.015 \leq z \leq 0.7026$, compiled in \cite{DES_2019d}. See \cite{Caoetal_2021b} for a description of how we use these SN Ia data.

The QSO data we use, that span the redshift range $0.462 \leq z \leq 2.73$, are listed in Table 1 of \cite{Cao_et_al2017b}. These consist of 120 measurements of the angular size
\begin{equation}
    \theta(z) = \frac{l_{\rm m}}{D_{A}(z)}.
\end{equation}
Here $l_{\rm m}$ is the characteristic linear size of QSOs in the sample and $D_{A}$ (defined below) is the angular size distance. Here we improve on the approach of \cite{Cao_et_al2017b}, \cite{Ryanetal2019}, and \cite{Caoetal_2020, Caoetal_2021a, Caoetal_2021b}, by treating $l_{\rm m}$ as a nuisance parameter to be determined from these measurements so that these QSO data are independent of $H(z)$ data.

The old \hiig\ data (which we dub ``\hiig-2019'') consist of 107 low redshift measurements that span $0.0088 \leq z \leq 0.16417$, used in \cite{Chavez_2014} (recalibrated by \citealp{G-M_2019}), and 46 high redshift measurements that span $0.636427 \leq z \leq 2.42935$. The new \hiig-2021 data, comprising the original 107 low redshift measurements and 74 updated high redshift measurements (that now span $0.636427 \leq z \leq 2.545$), are listed in Table A3 of \cite{GM2021}. 

The correlation between \hiig\ luminosity ($L$) and velocity dispersion ($\sigma$) is
\begin{equation}
\label{eq:logL}
    \log L = \beta \log \sigma + \gamma,
\end{equation}
where $\beta$ and $\gamma$ are the slope and intercept. $\log = \log_{10}$ is implied everywhere. Both \hiig\ data sets are corrected for extinction by using the \cite{Gordon_2003} extinction law, with
\begin{equation}
    \label{eq:Gordon_beta}
    \beta = 5.022 \pm 0.058,
\end{equation}
and
\begin{equation}
    \label{eq:Gordon_gamma}
    \gamma = 33.268 \pm 0.083.
\end{equation}
A detailed description of how to use \hiig\ data can be found in \cite{Caoetal_2020}. Note that the systematic uncertainties of both \hiig\ and QSO data are not considered so that the reduced $\chi^2$'s are relatively large.

The transverse comoving distance $D_M(z)$, the luminosity distance $D_L(z)$, and the angular size distance $D_A(z)$ are related through $D_M(z)=D_L(z)/(1+z)=(1+z)D_A(z)$.
\be
\label{eq:DL}
\resizebox{0.45\textwidth}{!}{%
  $D_M(z) = 
    \begin{cases}
    \vspace{1mm}
    D_C(z) & \text{if}\ \Omega_{\rm k_0} = 0,\\
    \vspace{1mm}
    \frac{c}{H_0\sqrt{\Omega_{\rm k_0}}}\sinh\left[\sqrt{\Omega_{\rm k_0}}H_0D_C(z)/c\right] & \text{if}\ \Omega_{\rm k_0} > 0, \\
    \vspace{1mm}
    \frac{c}{H_0\sqrt{|\Omega_{\rm k_0}|}}\sin\left[\sqrt{|\Omega_{\rm k_0}|}H_0D_C(z)/c\right] & \text{if}\ \Omega_{\rm k_0} < 0,
    \end{cases}$%
    }
\ee
where
\be
\label{eq:DC}
    D_C(z) \equiv c\int^z_0 \frac{dz'}{H(z')},
\ee
with $c$ being the speed of light (\citealp{Hogg}).

\section{Data Analysis Methodology}
\label{sec:analysis}

In this paper we use the \textsc{class} code to compute cosmological model predictions as a function of the cosmological model and other parameters. These predictions are compared to observational data using the Markov chain Monte Carlo (MCMC) code \textsc{MontePython} \citep{Audren:2012wb} to maximize the likelihood function, $\mathcal{L}$, and thereby determine the best-fitting values of the free parameters. The priors on the cosmological parameters are flat and nonzero over the same ranges as used in \cite{Caoetal_2021b}, except that now $\obh\in[0.00499, 0.03993]$.\footnote{The value of primordial Helium abundance $Y_{\rm p}$ is set using a standard big-bang nucleosynthesis prediction by interpolation on a grid of values computed using version 1.2 of the PArthENoPE BBN code for a neutron lifetime of 880.2 s. Since we choose the effective number of relativistic neutrino species $N_{\rm eff}=3.046$, \obhs\ is therefore limited to the range of [0.00499, 0.03993] by the correlated predictions of $Y_{\rm p}$.} The prior range of the QSO nuisance parameter $l_{\rm m}$ is not bounded.

The computation of the likelihood functions of $H(z)$, BAO, \hiig, and QSO data are described in \cite{Caoetal_2020} and \cite{Caoetal_2021a}, whereas that of the likelihood functions of SN Ia measurements can be found in \cite{Caoetal_2021b}. One can also find the definitions of the Akaike Information Criterion ($AIC$) and the Bayesian Information Criterion ($BIC$) in those papers.

\section{Results}
\label{sec:results}

\begin{figure*}
\centering
 \subfloat[]{%
    \includegraphics[width=3.5in,height=3.5in]{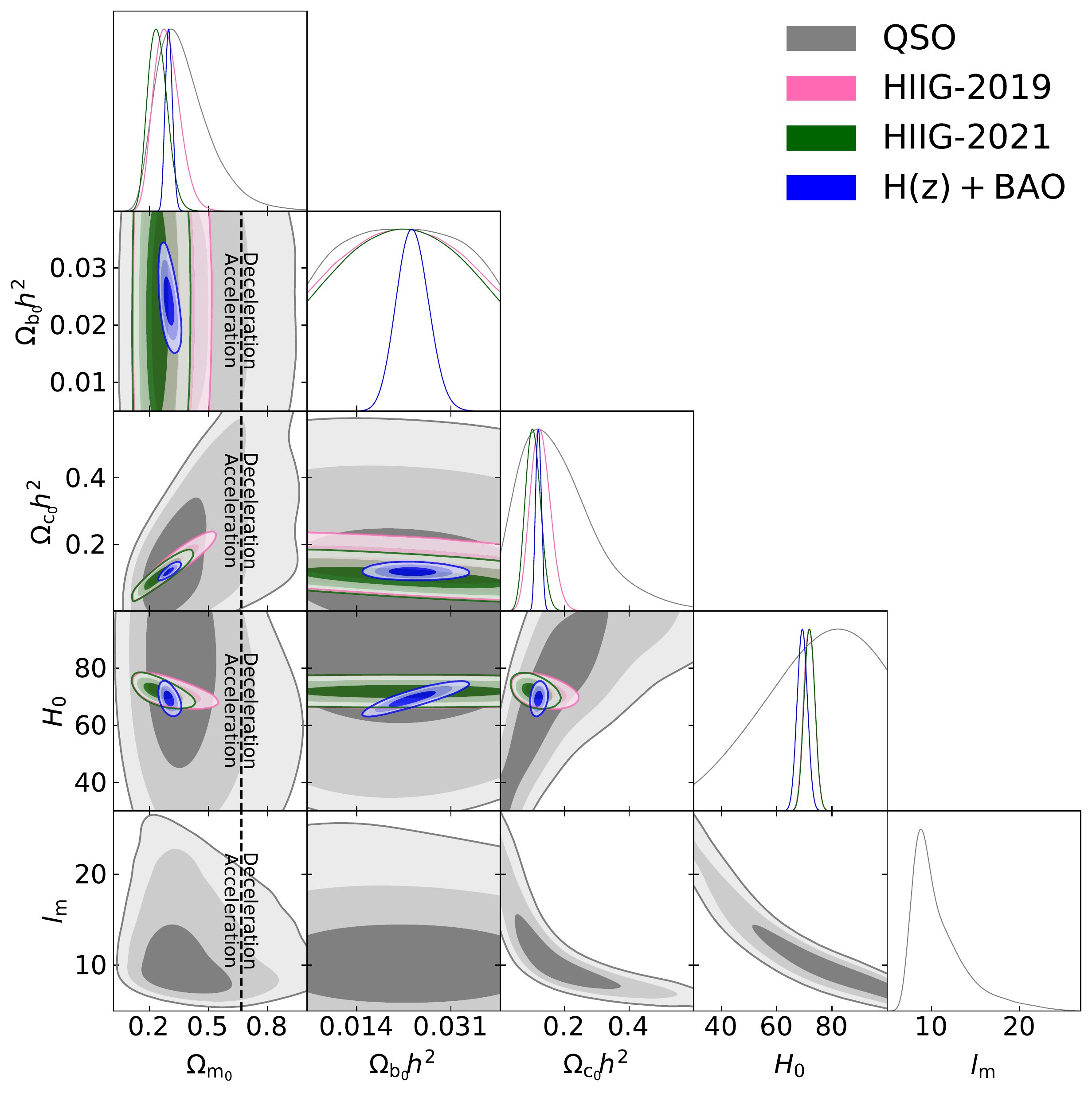}}
 \subfloat[]{%
    \includegraphics[width=3.5in,height=3.5in]{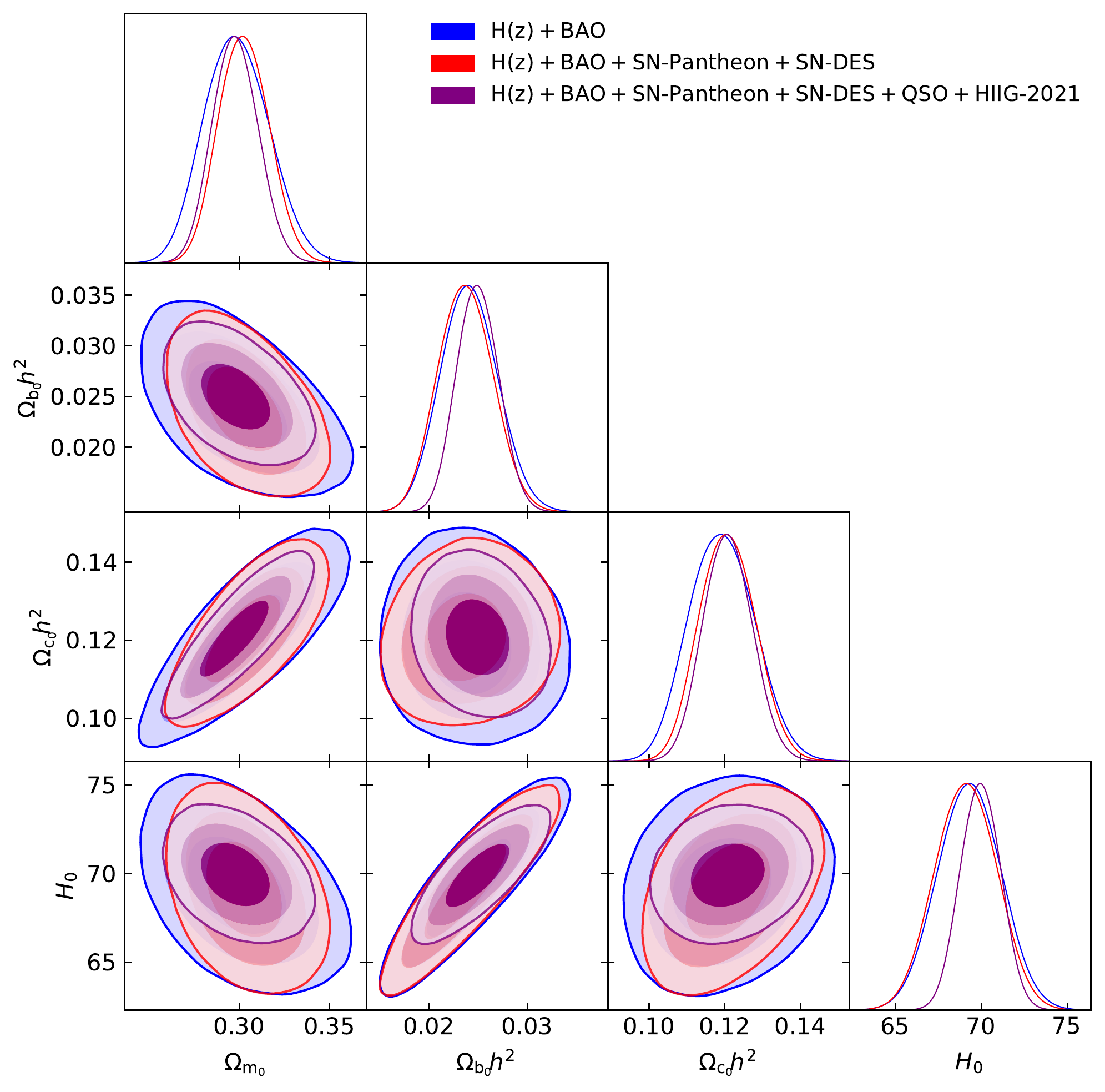}}\\
\caption{One-dimensional likelihoods and 1$\sigma$, 2$\sigma$, and 3$\sigma$ two-dimensional likelihood confidence contours for flat \lcdm. Left panel shows individual data set and $H(z)$ + BAO results and the right panel shows joint data sets constraints. The zero-acceleration lines are shown as black dashed lines in the left panel, which divide the parameter space into regions associated with currently-accelerating (left) and currently-decelerating (right) cosmological expansion, while in the right panel the joint analyses favor currently-accelerating expansion.}
\label{fig1}
\end{figure*}

\begin{figure*}
\centering
 \subfloat[]{%
    \includegraphics[width=3.5in,height=3.5in]{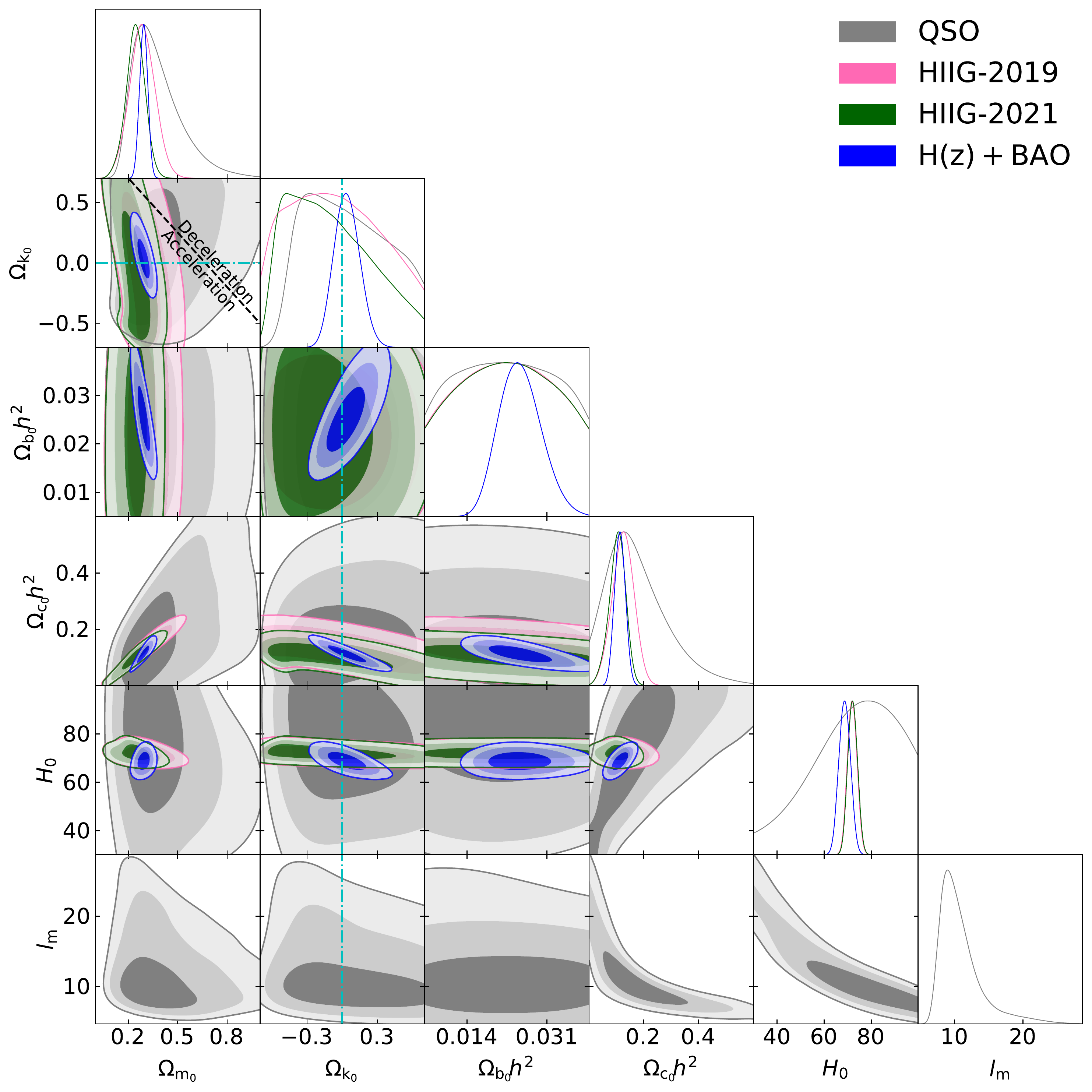}}
 \subfloat[]{%
    \includegraphics[width=3.5in,height=3.5in]{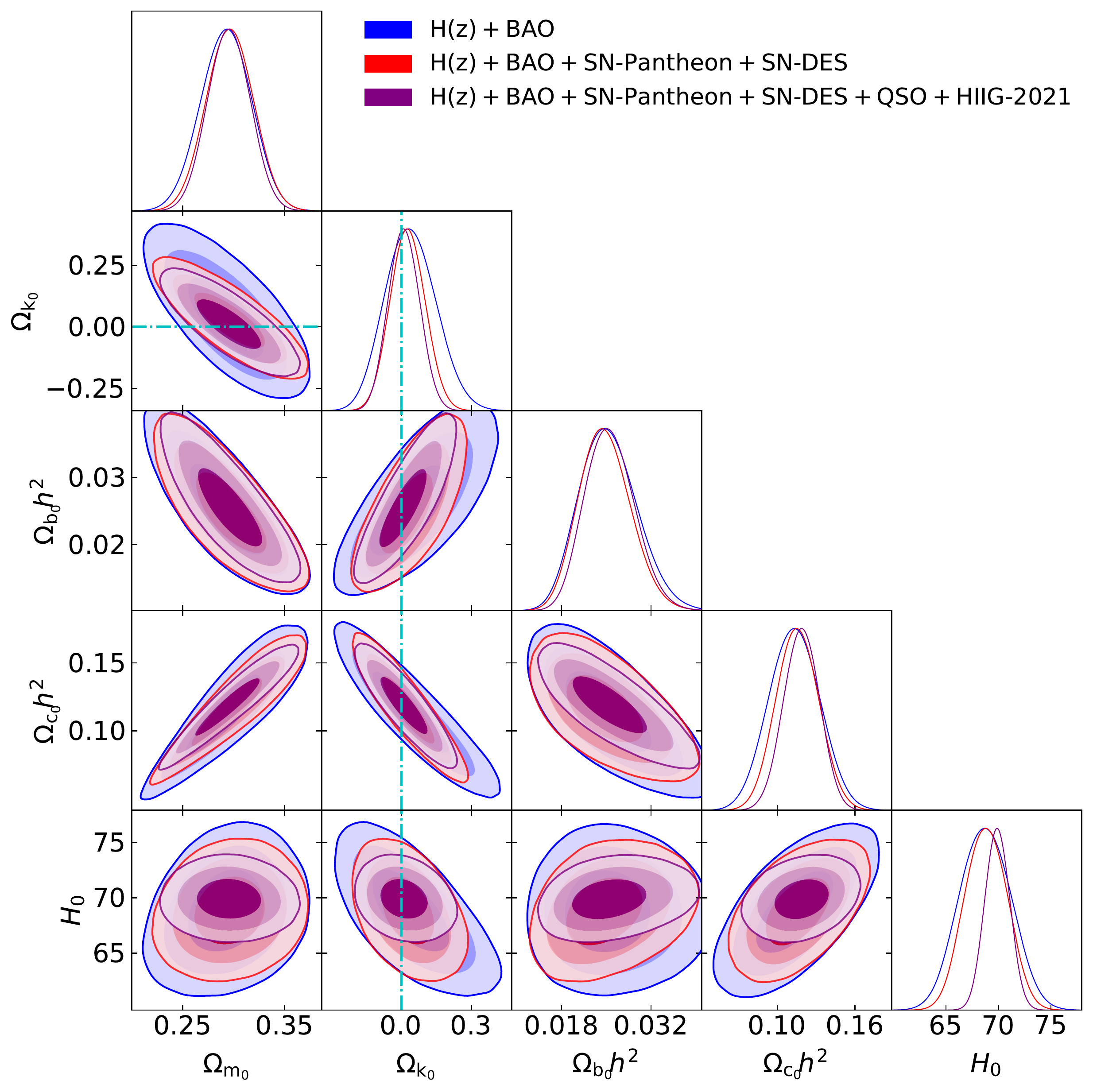}}\\
\caption{Same as Fig. \ref{fig1} but for non-flat \lcdm. The flat \lcdm\ case is shown as the cyan dash-dot lines, with closed spatial hypersurfaces either below or to the left. The black dashed line in the left panel is the zero-acceleration line, which divides the parameter space into regions associated with currently-accelerating (below left) and currently-decelerating (above right) cosmological expansion. In the right panel, the joint analyses favor currently-accelerating expansion.}
\label{fig2}
\end{figure*}

\begin{figure*}
\centering
 \subfloat[]{%
    \includegraphics[width=3.5in,height=3.5in]{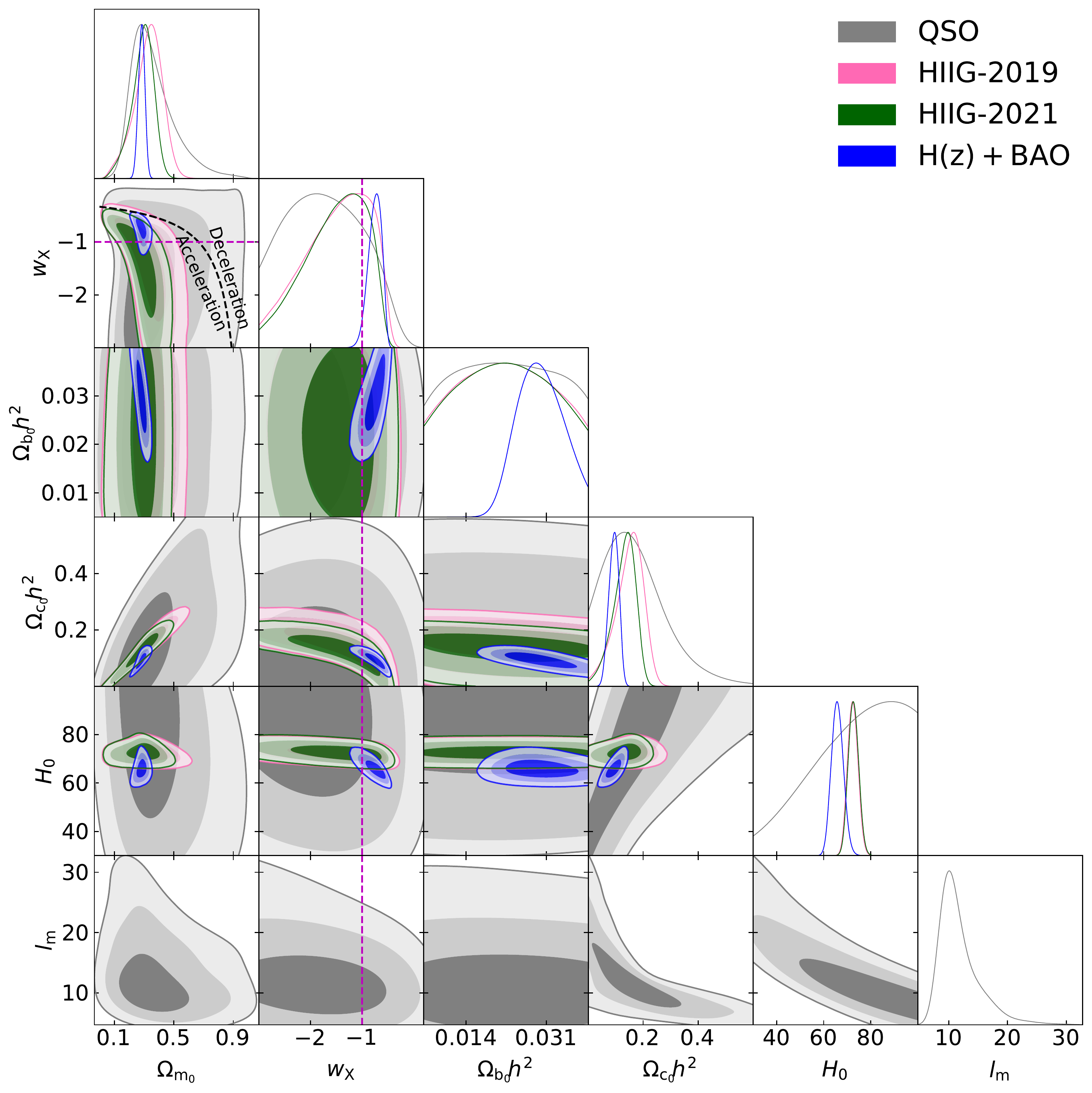}}
 \subfloat[]{%
    \includegraphics[width=3.5in,height=3.5in]{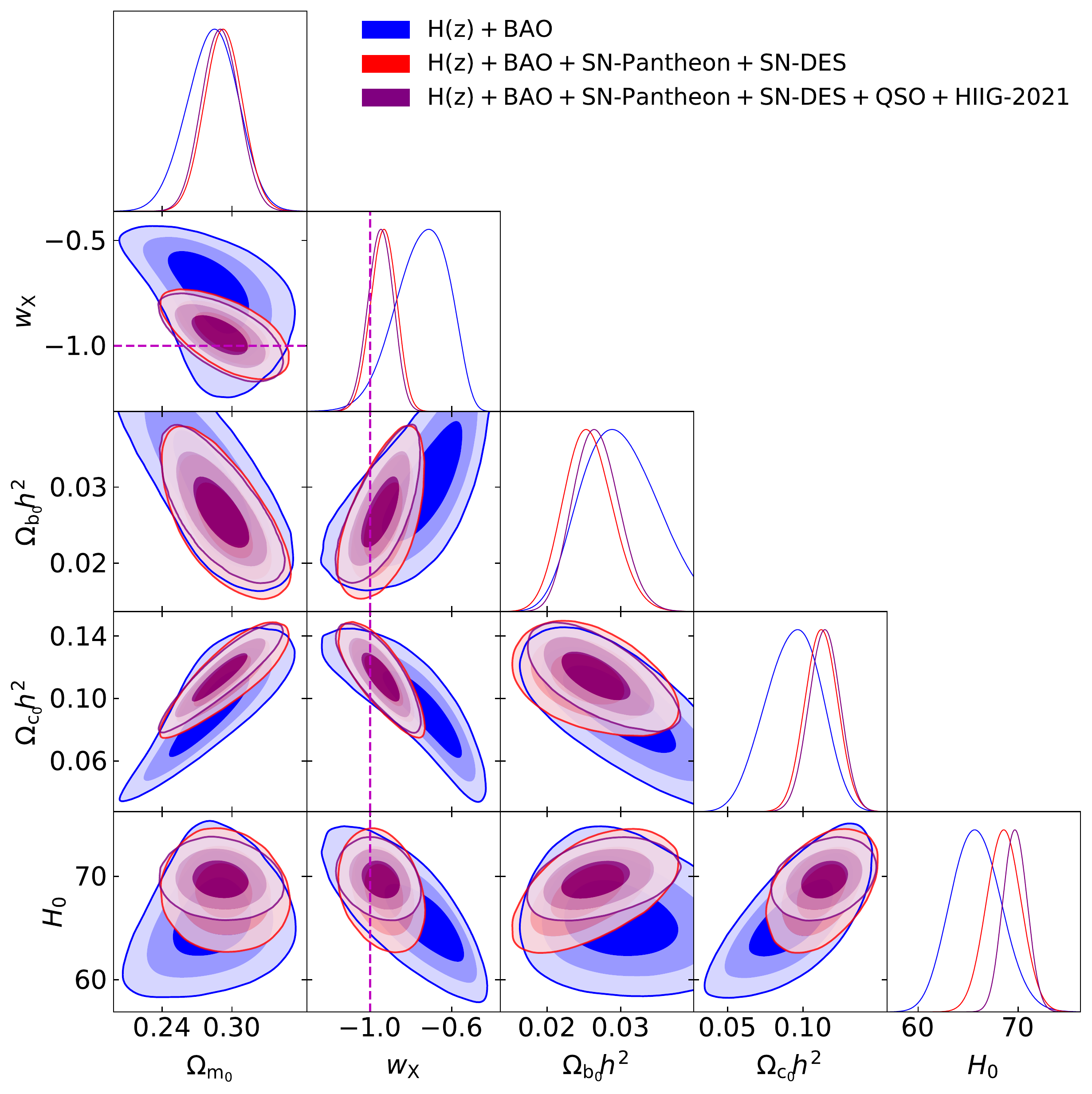}}\\
\caption{One-dimensional likelihoods and 1$\sigma$, 2$\sigma$, and 3$\sigma$ two-dimensional likelihood confidence contours for flat XCDM. The black dashed line in the left panel is the zero-acceleration line, which divides the parameter space into regions associated with currently-accelerating (below left) and currently-decelerating (above right) cosmological expansion. In the right panel, the joint analyses favor currently-accelerating expansion. The magenta lines represent $w_{\rm X}=-1$, i.e. the flat \lcdm\ model.}
\label{fig3}
\end{figure*}

\begin{figure*}
\centering
 \subfloat[]{%
    \includegraphics[width=3.5in,height=3.5in]{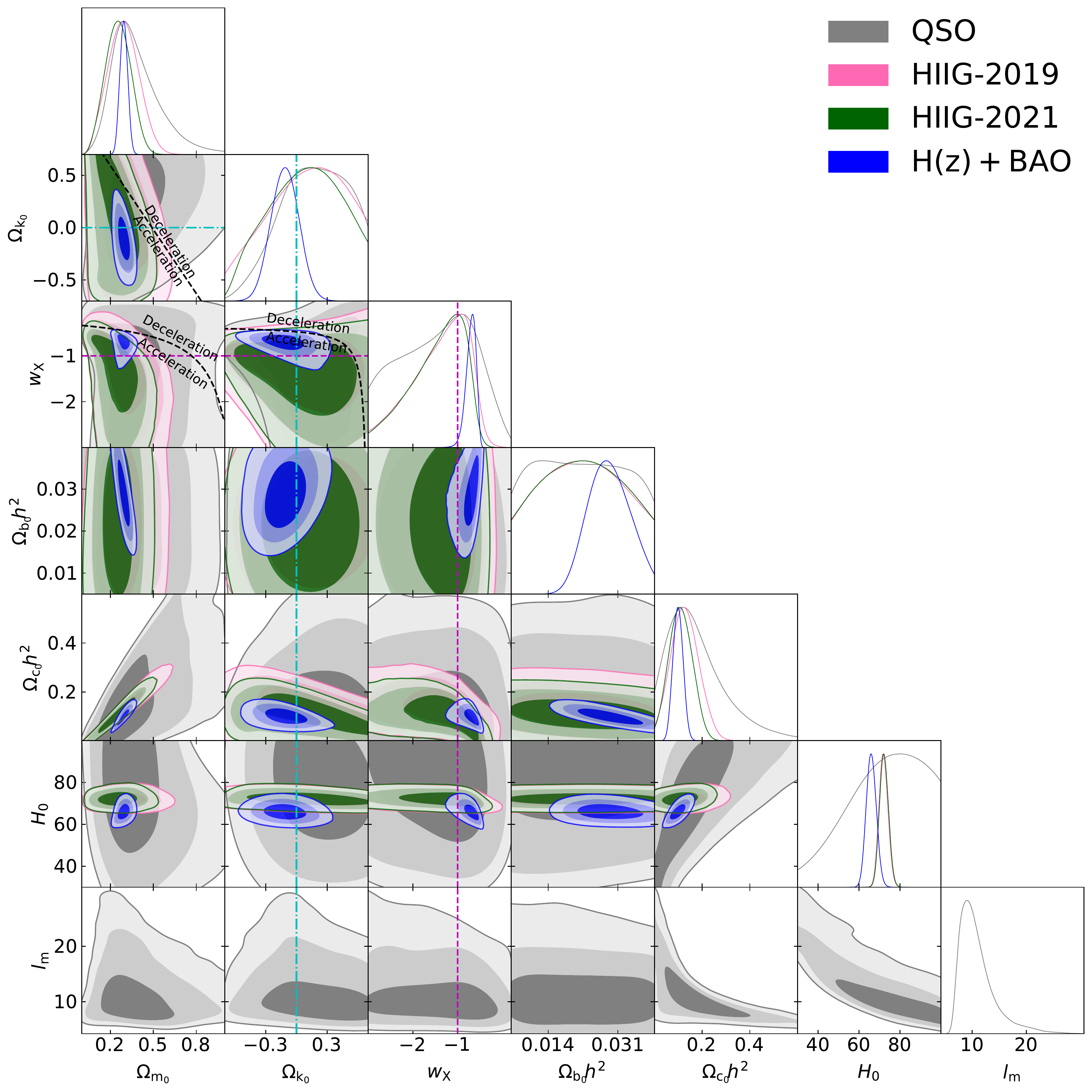}}
 \subfloat[]{%
    \includegraphics[width=3.5in,height=3.5in]{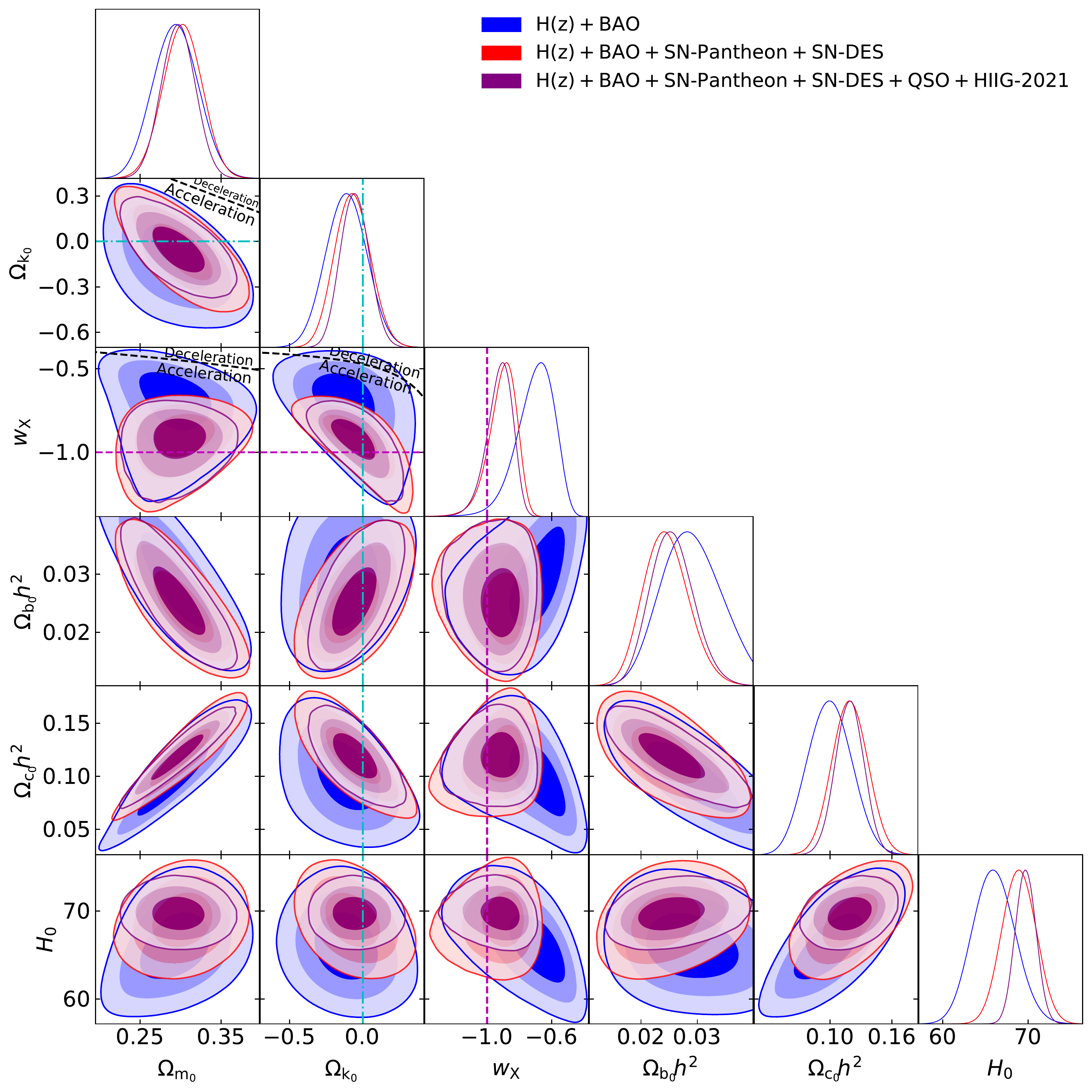}}\\
\caption{Same as Fig. \ref{fig3} but for non-flat XCDM, where the black dashed zero-acceleration lines are computed for the third cosmological parameter set to the $H(z)$ + BAO data best-fitting values listed in Table \ref{tab:BFP}, with currently-accelerating cosmological expansion residing below left. The flat XCDM case is denoted as the cyan dash-dot lines, with closed spatial hypersurfaces either below or to the left. The magenta lines represent $w_{\rm X} = -1$, i.e. the non-flat \lcdm\ model. In all cases except for the QSO only case, almost all of the favored parameter space is associated with currently-accelerating cosmological expansion.}
\label{fig4}
\end{figure*}

\begin{figure*}
\centering
 \subfloat[]{%
    \includegraphics[width=3.5in,height=3.5in]{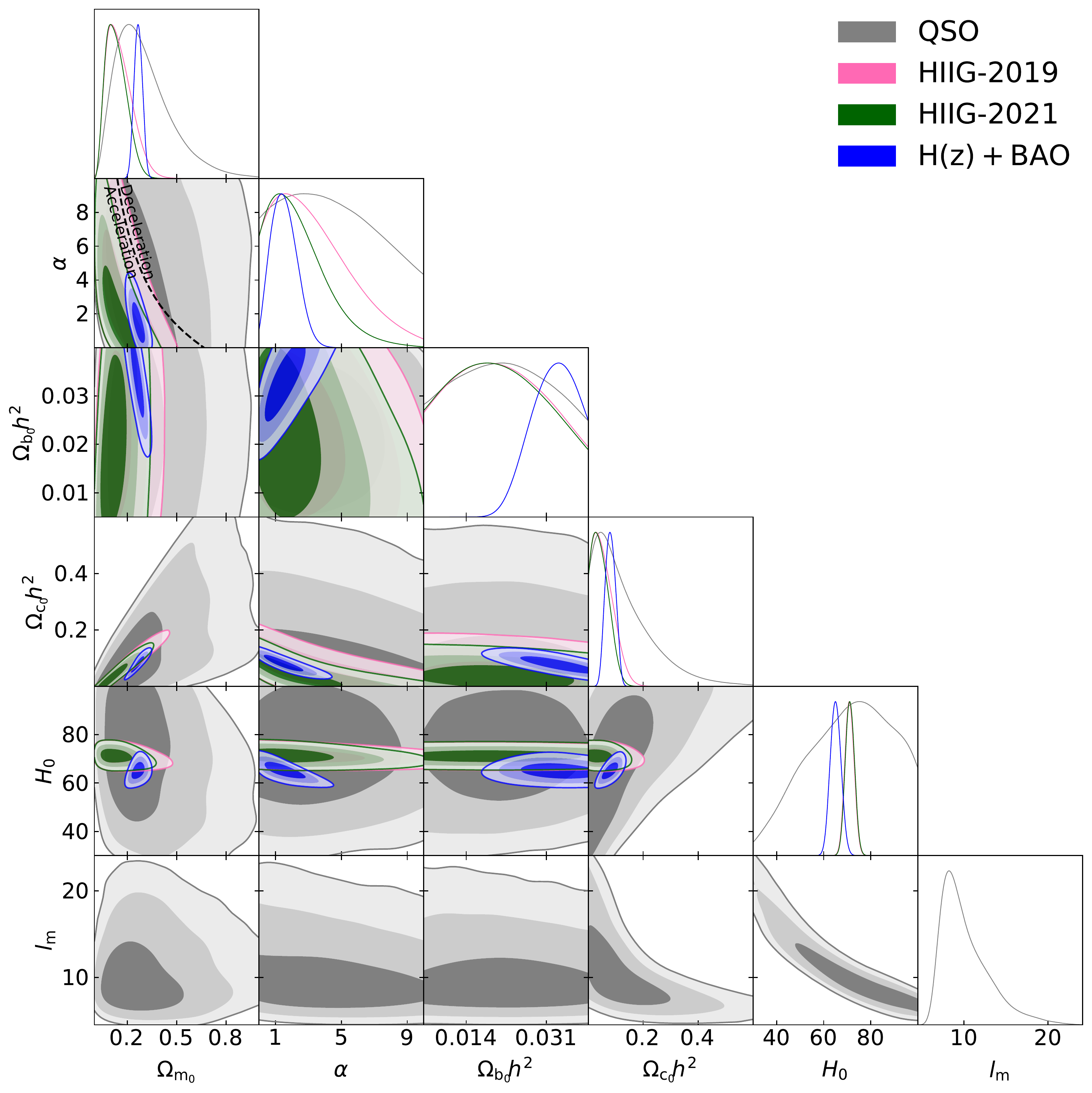}}
 \subfloat[]{%
    \includegraphics[width=3.5in,height=3.5in]{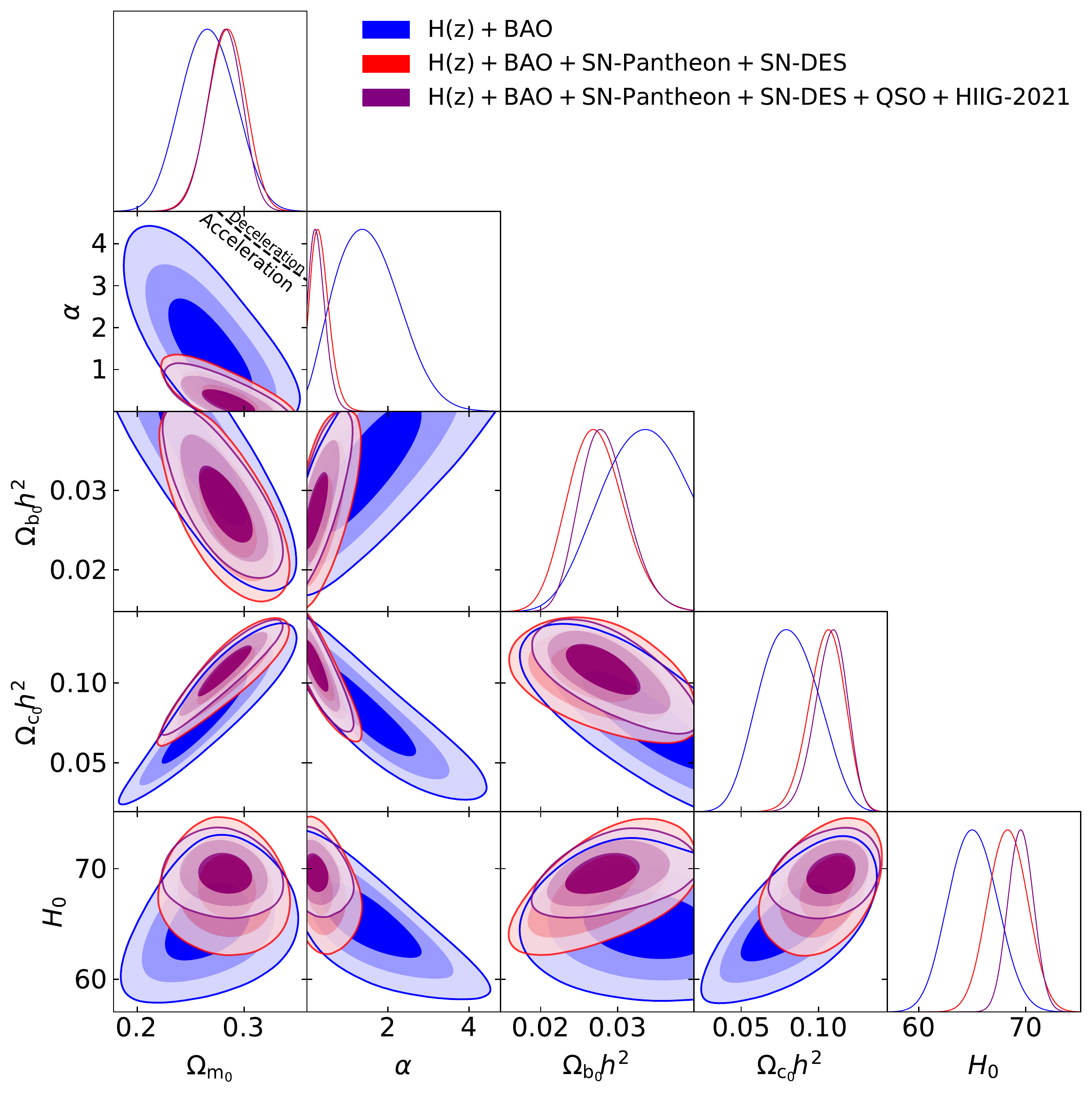}}\\
\caption{One-dimensional likelihoods and 1$\sigma$, 2$\sigma$, and 3$\sigma$ two-dimensional likelihood confidence contours for flat \pcdm. The black dashed lines are the zero-acceleration lines, which divides the parameter space into regions associated with currently-accelerating (below left) and currently-decelerating (above right) cosmological expansion. The $\alpha = 0$ axis is the flat \lcdm\ model. In all cases except for the QSO only case, almost all of the favored parameter space is associated with currently-accelerating cosmological expansion.}
\label{fig5}
\end{figure*}

\begin{figure*}
\centering
 \subfloat[]{%
    \includegraphics[width=3.5in,height=3.5in]{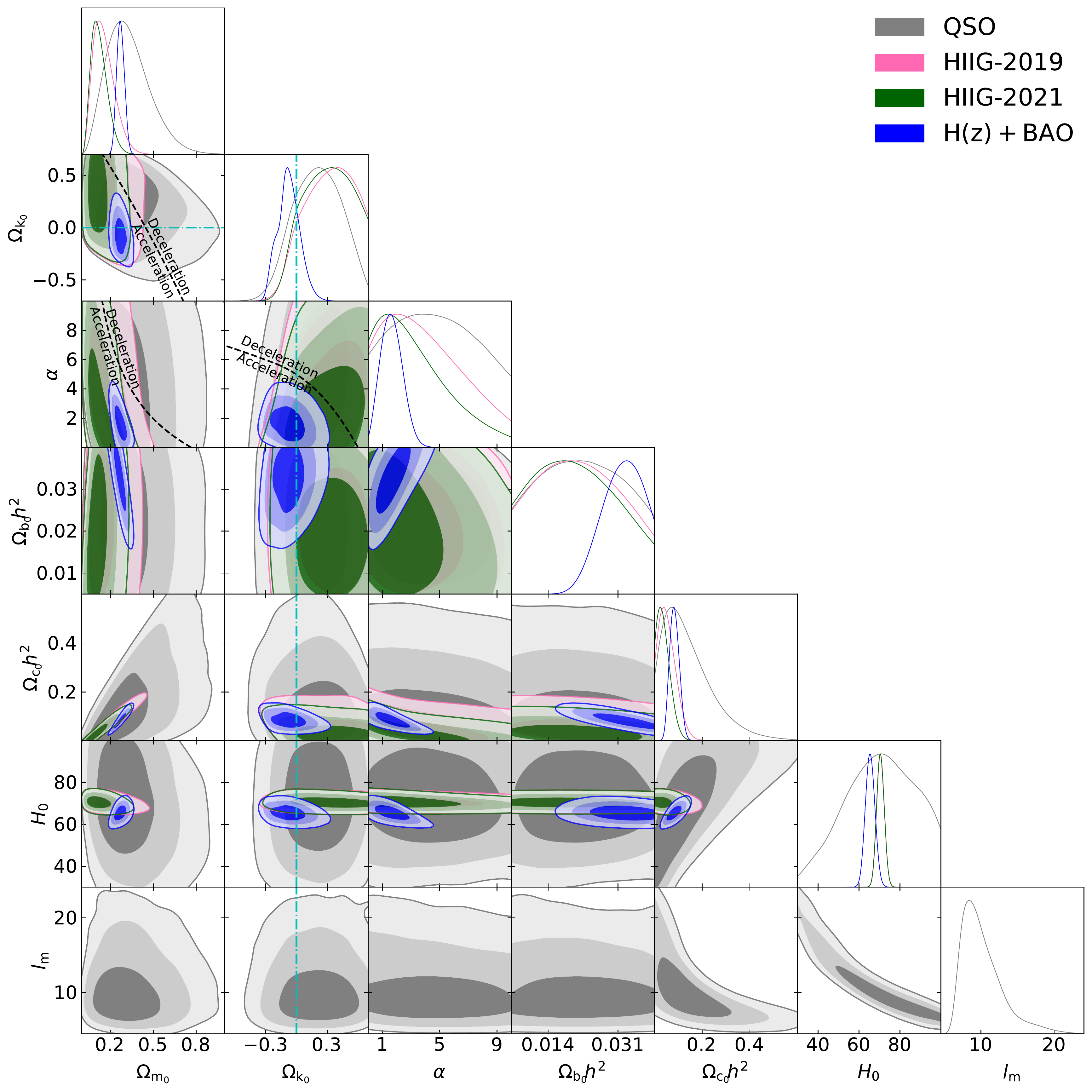}}
 \subfloat[]{%
    \includegraphics[width=3.5in,height=3.5in]{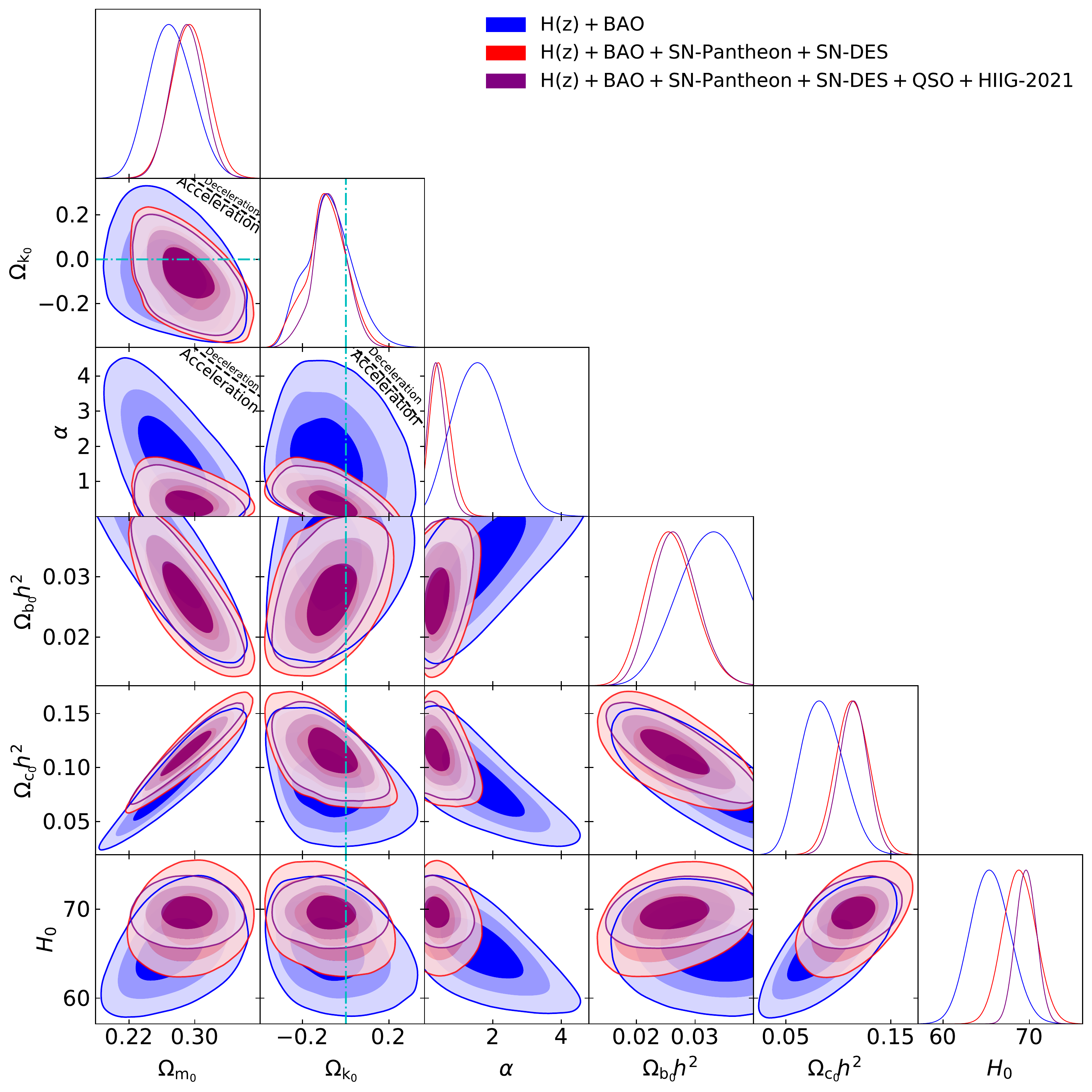}}\\
\caption{Same as Fig. \ref{fig5} but for non-flat \pcdm, where the black dashed zero-acceleration lines are computed for the third cosmological parameter set to the $H(z)$ + BAO data best-fitting values listed in Table \ref{tab:BFP}. Currently-accelerating cosmological expansion occurs below left of these lines. The cyan dash-dot lines represent the flat \pcdm\ case, with closed spatial geometry either below or to the left. The $\alpha = 0$ axis is the non-flat \lcdm\ model. In the right panel, the joint analyses favor currently-accelerating expansion. }
\label{fig6}
\end{figure*}

The posterior one-dimensional (1D) probability distributions and two-dimensional (2D) confidence regions of the cosmological and nuisance parameters for the six flat and non-flat models are shown in Figs. \ref{fig1}--\ref{fig6}, in gray (QSO), pink (\hiig-2019), green (\hiig-2021), blue ($H(z)$ + BAO), red ($H(z)$ + BAO + SN-Pantheon + SN-DES, HzBSNPD), and purple ($H(z)$ + BAO + SN-Pantheon + SN-DES + QSO + \hiig-2021, HzBSNPDQH). We list the unmarginalized best-fitting parameter values, as well as the corresponding $\chi^2$, $AIC$, $BIC$, degrees of freedom $\nu$ ($\nu \equiv N - n$), reduced $\chi^2$ ($\chi^2/\nu$), $\Delta \chi^2$, $\Delta AIC$, and $\Delta BIC$ for all models and data combinations, in Table \ref{tab:BFP}. The marginalized posterior mean parameter values and uncertainties ($\pm 1\sigma$ error bars or $2\sigma$ limits), for all models and data combinations, are listed in Table \ref{tab:1d_BFP}.\footnote{We use the \textsc{python} package \textsc{getdist} \citep{Lewis_2019} to determine the posterior means and uncertainties and to generate the marginalized likelihood contours.}

\begin{sidewaystable*}
\centering
\resizebox{1.0\columnwidth}{!}{%
\begin{threeparttable}
\caption{Unmarginalized best-fitting parameter values for all models from various combinations of data.}\label{tab:BFP}
\begin{tabular}{lccccccccccccccccc}
\toprule
Model & Data set & $\Omega_{\mathrm{b_0}}\!h^2$ & $\Omega_{\mathrm{c_0}}\!h^2$ & $\Omega_{\mathrm{m_0}}$ & $\Omega_{\mathrm{k_0}}$ & $w_{\mathrm{X}}$ & $\alpha$ & $H_0$\tnote{a} & $l_{\mathrm{m}}$\tnote{b} & $\chi^2$ & $\nu$ & $AIC$ & $BIC$ & $\chi^2/\nu$ & $\Delta \chi^2$ & $\Delta AIC$ & $\Delta BIC$ \\
\midrule
Flat \lcdm & $H(z)$ + BAO & 0.0239 & 0.1187 & 0.298 & -- & -- & -- & 69.13 & -- & 23.66 & 39 & 29.66 & 34.87 & 0.61 & 0.00 & 0.00 & 0.00\\
 & \hiig-2019 & 0.0200 & 0.1215 & 0.274 & -- & -- & -- & 71.86 & -- & 410.75 & 150 & 416.75 & 425.84 & 2.74 & 0.00 & 0.00 & 0.00\\
 & \hiig-2021 & 0.0156 & 0.1058 & 0.235 & -- & -- & -- & 71.89 & -- & 433.86 & 178 & 439.86 & 449.45 & 2.44 & 0.00 & 0.00 & 0.00\\
 & QSO & 0.0095 & 0.0172 & 0.315 & -- & -- & -- & 29.14 & 25.99 & 352.04 & 116 & 360.04 & 371.19 & 3.03 & 0.00 & 0.00 & 0.00\\
 & HzBSNPD\tnote{c} & 0.0235 & 0.1200 & 0.302 & -- & -- & -- & 68.90 & -- & 1080.48 & 1107 & 1086.48 & 1101.52 & 0.98 & 0.00 & 0.00 & 0.00\\
 & HzBSNPDQH\tnote{d} & 0.0248 & 0.1205 & 0.298 & -- & -- & -- & 69.83 & 10.98 & 1868.64 & 1407 & 1876.64 & 1897.64 & 1.33 & 0.00 & 0.00 & 0.00\\
\\
Non-flat \lcdm & $H(z)$ + BAO & 0.0247 & 0.1140 & 0.294 & 0.029 & -- & -- & 68.68 & -- & 23.60 & 38 & 31.60 & 38.55 & 0.62 & $-0.06$ & 1.94 & 3.68\\
 & \hiig-2019 & 0.0075 & 0.1585 & 0.314 & $-0.424$ & -- & -- & 72.70 & -- & 410.40 & 149 & 418.40 & 430.52 & 2.75 & $-0.35$ & 1.65 & 4.68\\
 & \hiig-2021 & 0.0114 & 0.1282 & 0.260 & $-0.490$ & -- & -- & 73.20 & -- & 432.79 & 177 & 440.79 & 453.58 & 2.45 & $-1.07$ & 0.93 & 4.13\\
 & QSO & 0.0320 & 0.1277 & 0.236 & $-0.363$ & -- & -- & 82.31 & 10.52 & 351.12 & 115 & 361.12 & 375.06 & 3.05 & $-0.92$ & 1.08 & 3.87\\
 & HzBSNPD\tnote{c} & 0.0243 & 0.1153 & 0.296 & 0.025 & -- & -- & 68.61 & -- & 1080.38 & 1106 & 1088.38 & 1108.42 & 0.98 & $-0.10$ & 1.90 & 6.90\\
 & HzBSNPDQH\tnote{d} & 0.0249 & 0.1192 & 0.296 & 0.005 & -- & -- & 69.76 & 10.99 & 1868.63 & 1406 & 1878.63 & 1904.89 & 1.33 & $-0.01$ & 1.99 & 7.25\\
\\
Flat XCDM & $H(z)$ + BAO & 0.0304 & 0.0891 & 0.281 & -- & $-0.701$ & -- & 65.18 & -- & 19.65 & 38 & 27.65 & 34.60 & 0.52 & $-4.01$ & $-2.01$ & $-0.27$\\
 & \hiig-2019 & 0.0107 & 0.1180 & 0.251 & -- & $-0.896$ & -- & 71.62 & -- & 410.72 & 149 & 418.72 & 430.85 & 2.76 & $-0.03$ & 1.97 & 5.01\\
 & \hiig-2021 & 0.0291 & 0.0942 & 0.239 & -- & $-1.013$ & -- & 71.88 & -- & 433.86 & 177 & 441.86 & 454.65 & 2.45 & 0.00 & 2.00 & 5.20\\
 & QSO & 0.0173 & 0.0050 & 0.253 & -- & $-2.137$ & -- & 29.68 & 31.24 & 351.83 & 115 & 361.83 & 375.77 & 3.06 & $-0.21$ & 1.79 & 4.58\\
 & HzBSNPD\tnote{c} & 0.0254 & 0.1119 & 0.293 & -- & $-0.934$ & -- & 68.51 & -- & 1079.24 & 1106 & 1087.24 & 1107.29 & 0.98 & $-1.24$ & 0.76 & 5.77\\
 & HzBSNPDQH\tnote{d} & 0.0266 & 0.1135 & 0.289 & -- & $-0.948$ & -- & 69.63 & 10.95 & 1867.85 & 1406 & 1877.85 & 1904.11 & 1.33 & $-0.79$ & 1.21 & 6.47\\
\\
Non-flat XCDM & $H(z)$ + BAO & 0.0290 & 0.0980 & 0.295 & $-0.152$ & $-0.655$ & -- & 65.59 & -- & 18.31 & 37 & 28.31 & 37.00 & 0.49 & $-5.35$ & $-1.35$ & 2.13\\
 & \hiig-2019 & 0.0128 & 0.0011 & 0.027 & $-0.625$ & $-0.630$ & -- & 72.24 & -- & 405.56 & 148 & 415.56 & 430.72 & 2.74 & $-5.19$ & $-1.19$ & 4.88\\
 & \hiig-2021 & 0.0260 & 0.0066 & 0.062 & $-0.573$ & $-0.680$ & -- & 72.43 & -- & 430.07 & 176 & 440.07 & 456.06 & 2.44 & $-3.79$ & 0.21 & 6.61\\
 & QSO & 0.0153 & 0.0122 & 0.060 & $-0.570$ & $-0.617$ & -- & 67.92 & 11.82 & 350.22 & 114 & 362.22 & 378.95 & 3.07 & $-1.82$ & 2.18 & 7.76\\
 & HzBSNPD\tnote{c} & 0.0234 & 0.1232 & 0.309 & $-0.111$ & $-0.876$ & -- & 68.94 & -- & 1078.38 & 1105 & 1088.38 & 1113.44 & 0.98 & $-2.10$ & 1.90 & 11.92\\
 & HzBSNPDQH\tnote{d} & 0.0246 & 0.1208 & 0.299 & $-0.083$ & $-0.900$ & -- & 69.72 & 10.89 & 1867.11 & 1405 & 1879.11 & 1910.63 & 1.33 & $-1.53$ & 2.47 & 12.99\\
\\
Flat $\phi$CDM & $H(z)$ + BAO & 0.0333 & 0.0788 & 0.264 & -- & -- & 1.504 & 65.20 & -- & 19.49 & 38 & 27.49 & 34.44 & 0.51 & $-4.17$ & $-2.17$ & $-0.43$\\
 & \hiig-2019 & 0.0304 & 0.1038 & 0.261 & -- & -- & 0.174 & 71.75 & -- & 410.74 & 149 & 418.74 & 430.86 & 2.76 & $-0.01$ & 1.99 & 5.02\\
 & \hiig-2021 & 0.0334 & 0.0876 & 0.234 & -- & -- & 0.001 & 71.94 & -- & 433.86 & 177 & 441.86 & 454.65 & 2.45 & 0.00 & 2.00 & 5.20\\
 & QSO & 0.0370 & 0.0960 & 0.316 & -- & -- & 0.001 & 64.91 & 11.66 & 352.05 & 115 & 362.05 & 375.98 & 3.06 & 0.01 & 2.01 & 4.79\\
 & HzBSNPD\tnote{c} & 0.0257 & 0.1097 & 0.290 & -- & -- & 0.226 & 68.38 & -- & 1079.09 & 1106 & 1087.09 & 1107.14 & 0.98 & $-1.39$ & 0.61 & 5.62\\
 & HzBSNPDQH\tnote{d} & 0.0270 & 0.1125 & 0.288 & -- & -- & 0.174 & 69.63 & 10.94 & 1867.75 & 1406 & 1877.75 & 1904.01 & 1.33 & $-0.89$ & 1.11 & 6.37\\
\\
Non-flat $\phi$CDM & $H(z)$ + BAO & 0.0334 & 0.0816 & 0.266 & $-0.147$ & -- & 1.915 & 65.70 & -- & 18.15 & 37 & 28.15 & 36.84 & 0.49 & $-5.51$ & $-1.51$ & 1.97\\
 & \hiig-2019 & 0.0251 & 0.1126 & 0.265 & $-0.265$ & -- & 0.433 & 72.04 & -- & 410.37 & 148 & 420.37 & 435.52 & 2.77 & $-0.38$ & 3.62 & 9.68\\
 & \hiig-2021 & 0.0146 & 0.1014 & 0.249 & $-0.246$ & -- & 0.101 & 72.48 & -- & 433.19 & 176 & 443.19 & 459.19 & 2.46 & $-0.67$ & 3.33 & 9.74\\
 & QSO & 0.0282 & 0.0469 & 0.261 & $-0.261$ & -- & 0.008 & 53.61 & 15.39 & 351.32 & 114 & 363.32 & 380.04 & 3.08 & $-0.72$ & 3.28 & 8.85\\
 & HzBSNPD\tnote{c} & 0.0243 & 0.1197 & 0.302 & $-0.110$ & -- & 0.442 & 69.05 & -- & 1078.07 & 1105 & 1088.07 & 1113.13 & 0.98 & $-2.41$ & 1.59 & 11.61\\
 & HzBSNPDQH\tnote{d} & 0.0250 & 0.1192 & 0.297 & $-0.092$ & -- & 0.348 & 69.69 & 10.87 & 1866.93 & 1405 & 1878.93 & 1910.45 & 1.33 & $-1.71$ & 2.29 & 12.81\\
\bottomrule
\end{tabular}%}
\begin{tablenotes}[flushleft]
\item [a] \hunit.
\item [b] pc.
\item [c] $H(z)$ + BAO + SN-Pantheon + SN-DES.
\item [d] $H(z)$ + BAO + SN-Pantheon + SN-DES + QSO + \hiig-2021.
\end{tablenotes}
\end{threeparttable}%
}
\end{sidewaystable*}

\subsection{Constraints from \hiig-2021 versus \hiig-2019}
\label{subsec:Results_HIIG_comparison}

Here we compare constraints from the current compilation of \hiig\ data (\hiig-2021) with the constraints from the older compilation (\hiig-2019). 

For both data sets, from the figures, most of the probability lies in the part of parameter space that corresponds to currently accelerating cosmological expansion.

For both the flat and non-flat \lcdm\ models, \hiig-2021 data favor lower values of \ochs\ and \om. For the non-flat case, the \hiig-2021 data prefer a lower (more negative) value of \ok\ than the \hiig-2019 data. It is worth noting that $H_0$ is not particularly sensitive to the change in the \hiig\ data. It is also worth noting that the \hiig-2021 compilation provides constraints on \obhs\ in the flat case, in contrast to \hiig-2019, although these constraints are not as tight as those obtained from the other data combinations. \hiig-2021 data also more tightly constrain \ok\ compared to the \hiig-2019 data. In comparison with the flat \lcdm\ \hiig-2021 constraints given in \cite{GM2021}, $h=0.717\pm0.018$ and $\Om=0.243^{+0.039}_{-0.050}$, our constraints, $h=0.7191\pm0.0192$ and $\Om=0.243^{+0.039}_{-0.051}$, are a bit less restrictive (due to our models having more free parameters) but are consistent with theirs.

In the flat and non-flat XCDM parametrizations, the \hiig-2021 data favor lower values of \ochs\!, \om\!, \wx\!, and \ok\ than those favored by the \hiig-2019 data, while the constraints on \obhs\ and $H_0$ are consistent with each other. The flat XCDM \hiig-2021 constraints are $\{h, \Om, w_{\rm X}\}=\{0.719\pm0.020, 0.250^{+0.10}_{-0.061}, -1.19^{+0.46}_{-0.38}\}$ in \cite{GM2021}, whereas our results are $\{h, \Om, w_{\rm X}\}=\{0.7266\pm0.0219, 0.288^{+0.087}_{-0.058}, -1.527^{+0.786}_{-0.391}\}$ and consistent with theirs within 1$\sigma$, although due to different prior ranges the posterior means deviate more than those for the flat \lcdm\ model.

In the flat \pcdm\ model, the \hiig-2021 data prefer 
lower values of \ochs\!, \om\!, and $\alpha$, with consistent constraints on \obhs\ and $H_0$. The \hiig-2021 data constrain $\alpha$ more tightly than the \hiig-2019 data, leading to $\alpha$ being consistent with zero to within a little more than 1$\sigma$. In the non-flat case, the \hiig-2021 data prefer lower values for all parameters. It is worth noting that both \hiig-2021 and \hiig-2019 data in the flat and non-flat \pcdm\ models determine lower values of \om, and \hiig-2021 data prefer the lowest \om\ value in the non-flat \pcdm\ model among all models.

\hiig-2021 and \hiig-2019 data result in higher values of $H_0$ than the other probes we study in this paper. The highest $H_0$ values are in the flat XCDM parametrization and are $72.66\pm2.19$ \hunit and $72.37^{+2.18}_{-2.20}$ \hunit, respectively, which are $1.31\sigma$ and $1.23\sigma$ higher than the median statistics estimate of $H_0=68 \pm 2.8$ \hunit\ \citep{chenratmed}, and $0.21\sigma$ and $0.33\sigma$ lower than the local Hubble constant measurement of $H_0 = 73.2 \pm 1.3$ \hunit\ \citep{Riess_2021}. The lowest $H_0$ estimates are in the non-flat \pcdm\ model and are $70.49\pm1.81$ \hunit and $70.53\pm1.79$ \hunit, which are $0.75\sigma$ and $0.76\sigma$ higher than the median statistics estimate of $H_0=68 \pm 2.8$ \hunit, and $1.22\sigma$ and $1.21\sigma$ lower than the local Hubble constant measurement of $H_0 = 73.2 \pm 1.3$ \hunit.

In the non-flat \lcdm\ model, \hiig-2021 and \hiig-2019 data favor closed spatial hypersurfaces, while in the non-flat XCDM parametrization and the non-flat \pcdm\ model, they favor open spatial hypersurfaces. Only in the non-flat \pcdm\ model, however, is \ok\ more than 1$\sigma$ away from spatial flatness. Dark energy dynamics is favored by both data sets, but dark energy being a cosmological constant is not disfavored (it is within 1$\sigma$ or just a little bit more away).

\subsection{QSO constraints alone and in comparison to those from other probes}
\label{subsec:Results_QSO_constraints}

In this subsection we discuss the constraints we obtain solely from the QSO data. As mentioned in Sec. \ref{sec:data}, in this paper we improve on earlier analyses of the QSO angular size data by now treating $l_{\rm m}$, the characteristic linear size of QSOs, as a nuisance parameter to be determined from the observational data. From QSO data alone, in Table \ref{tab:1d_BFP}, $l_{\rm m}$ ranges from a low of $10.26^{+1.24}_{-3.42}$ pc for the non-flat \pcdm\ model to a high of $11.90^{+1.52}_{-4.17}$ pc for the flat XCDM parametrization, differing by just 0.38$\sigma$. These values are consistent from model to model, largely justifying the use of QSOs as standard rulers, with $l_{\rm m}=11.03$ pc, the value we used in our previous studies (taken from \citealp{Cao_et_al2017b}). However, ignoring the dependence on cosmological model and the $l_{\rm m}$ errors, as we and others have previously done, results in mildly biased and somewhat more restrictive QSO angular size constraints than is warranted by data. These deficiencies are corrected in our improved analyses here.\footnote{When QSO data are combined with other probes, as in the HzBSNPDQH combination, the model-independence of $l_{\rm m}$ is evident and the determination here is consistent with $l_{\rm m}=11.03 \pm 0.25$ pc found by \cite{Cao_et_al2017b}.}  Additionally, we note that in Table \ref{tab:BFP}, the best-fitting values of $H_0$ in flat \lcdm\ and flat XCDM appear to be unreasonably low. This strange behavior is caused by the large values of $l_{\rm m}$, which push the $H_0$ values lower to obtain locally minimized $\chi^2$ values. Specifically, from the form of the model-predicted angular size of a quasar,
\begin{equation}
    \theta\left(z\right) \propto \frac{l_{\rm m} H_0}{d_{\rm M}(z)}
\end{equation}
(where $d_{\rm M}(z) := \frac{H_0}{c}D_{\rm M}(z)$ and suppressing irrelevant parameters), we can see that a large value of $l_{\rm m}$ requires a small value of $H_0$ in order to keep $\theta\left(z\right)$ constant. Since $l_{\rm m}$ has an unbounded prior range, it can roam over a larger region of parameter space than $H_0$. It therefore has the freedom to move into regions of parameter space where its value is unusually large; if this happens, then $H_0$ must be made small to compensate. This is only a partial answer, since it does not account for the variation of \ok\ (the effect of which is more complex, as \ok\ is coupled to the redshift $z$ through the function $d_{\rm M}(z)$), and so does not fully capture the behavior of $\theta(z)$ across all models, but it does give some insight into the apparently anomalously low values of $H_0$ that appear in some cases.\footnote{The relatively higher values of $H_0$ seen in the $\phi$CDM models pose an apparent challenge to this explanation, but here the best-fitting values of \obhs\ and \ochs\ need to be taken into account. In comparing, for example, the flat \lcdm\ model to the flat \pcdm\ model (both of which have nearly identical best-fitting values of \om), we can see that the flat \pcdm\ model has larger best-fitting values of both \obhs\ and \ochs. From the defining relationship $\Om = (\obh + \och)/{h^2}$, keeping \om\ constant requires \obhs\ + \ochs\ and $H_0$ to vary in tandem. If \obhs\ and \ochs\ both increase, as they do in going from flat \lcdm\ to flat \pcdm, then $H_0$ must also increase. This then has the effect of lowering $l_{\rm m}$ (all other parameters being held fixed).}

From the results listed in Table \ref{tab:1d_BFP}, we can draw the following conclusions. First, QSO data alone can only constrain the values of \obhs\ in the flat and non-flat \pcdm\ models. Second, QSO data alone prefer higher values of \om, which are consistent with almost all other probes except for the non-flat \pcdm\ \hiig-2021 case (the posterior mean values being 1.1$\sigma$ away from each other in this case). Furthermore, QSO data alone do not give tight constraints on $H_0$ or \ok. Although in each non-flat model open geometry is favored, given the large error bars, flat geometry is within 1$\sigma$.

QSO data favor higher central values of \ochs\ and \om, in both flat and non-flat \lcdm, compared to the central values favored by the other probes (although QSO constraints have wider error bars than the other constraints). QSO data only very weakly constrain the value of $H_0$ in the flat \lcdm\ model, while the fit of QSO data to the non-flat \lcdm\ model produces a tighter constraint whose central value is closer to that of the \hiig\ data and the local value favored by \cite{Riess_2021} (with wide error bars, however). In both the flat and non-flat cases, the marginalized values of $l_{\rm m}$ are close to the value obtained by \cite{Cao_et_al2017b}, with the central value in the flat \lcdm\ model here being only 0.02 pc away from that of \cite{Cao_et_al2017b} (with wider error bars than what they found). QSO data do not provide strong evidence for non-zero spatial curvature in the non-flat \lcdm\ model, as the marginalized posterior mean value of \ok\ is consistent with \ok $= 0$ to within 1$\sigma$.

When we look at the flat and non-flat XCDM parametrizations, we find that QSO data again favor somewhat large values of \ochs\ and \om\ (but, as with flat and non-flat \lcdm, these have wide error bars) and weak constraints on $H_0$. The central value of $H_0$ in the non-flat case is more consistent with \cite{Riess_2021} and with the values derived from the \hiig\ data. In both cases we find that the marginalized values of $l_{\rm m}$ are consistent with that of \cite{Cao_et_al2017b}. We also find that QSO data favor values of \wx\ that are in the phantom regime (consistent with the findings from the \hiig\ data). In the non-flat case, QSO data favor a relatively large and positive central value (0.170) for \ok, corresponding to a spatially open universe, but the error bars are wide enough that this result is still consistent with spatial flatness.

Both the flat and non-flat \pcdm\ models have central values of \obhs\ from QSO data that are similar to earlier findings (specifically, they are close to the values of \obhs\ obtained for the flat \lcdm\ and \pcdm\ models by \citealp{ParkRatra2018, ParkRatra2019a}). Both flat and non-flat \pcdm\ have relatively high central values of \ochs\ and \om\ (compared to the other probes), both favor similar large values of $\alpha$ (consistent with $\alpha = 0$, however, to within 1.15$\sigma$ and 1.29$\sigma$ in the flat and non-flat cases, respectively), and both show weak constraints on $H_0$. Both flat and non-flat \pcdm\ favor posterior mean values of $l_{\rm m}$ that are consistent to within 1$\sigma$ with the central value obtained by \cite{Cao_et_al2017b}. Like non-flat XCDM, non-flat \pcdm\ favors a relatively large and positive value of \ok, that is nevertheless consistent with spatial flatness to within 1$\sigma$.

\begin{sidewaystable*}
\centering
\resizebox{0.92\columnwidth}{!}{%
\begin{threeparttable}
\caption{One-dimensional posterior mean parameter values and uncertainties ($\pm 1\sigma$ error bars or $2\sigma$ limits) for all models from various combinations of data.}\label{tab:1d_BFP}
\begin{tabular}{lccccccccc}
\toprule
Model & Data set & $\Omega_{\mathrm{b_0}}\!h^2$ & $\Omega_{\mathrm{c_0}}\!h^2$ & $\Omega_{\mathrm{m_0}}$ & $\Omega_{\mathrm{k_0}}$ & $w_{\mathrm{X}}$ & $\alpha$ & $H_0$\tnote{a} & $l_{\mathrm{m}}$\tnote{b}\\
\midrule
Flat \lcdm & $H(z)$ + BAO & $0.0241\pm0.0029$ & $0.1193^{+0.0082}_{-0.0090}$ & $0.299^{+0.017}_{-0.019}$ & -- & -- & -- & $69.29^{+1.84}_{-1.85}$ & -- \\
 & \hiig-2019 & -- & $0.1258^{+0.0278}_{-0.0335}$ & $0.289^{+0.054}_{-0.074}$ & -- & -- & -- & $71.80\pm1.94$ & -- \\
 & \hiig-2021 & $0.0225\pm0.0108$ & $0.1023^{+0.0197}_{-0.0229}$ & $0.243^{+0.039}_{-0.051}$ & -- & -- & -- & $71.91\pm1.92$ & -- \\
 & QSO & -- & $0.1874^{+0.0592}_{-0.1595}$ & $0.387^{+0.078}_{-0.177}$ & -- & -- & -- & $>38.09$ & $11.05^{+1.10}_{-3.85}$ \\
 & HzBSNPD\tnote{c} & $0.0237\pm0.0028$ & $0.1208\pm0.0074$ & $0.303^{+0.013}_{-0.014}$ & -- & -- & -- & $69.10\pm1.80$ & -- \\
 & HzBSNPDQH\tnote{d} & $0.0250\pm0.0021$ & $0.1208\pm0.0064$ & $0.298\pm0.013$ & -- & -- & -- & $69.95\pm1.18$ & $10.96\pm0.26$ \\
\\
Non-flat \lcdm & $H(z)$ + BAO & $0.0253^{+0.0040}_{-0.0049}$ & $0.1134^{+0.0196}_{-0.0197}$ & $0.293\pm0.025$ & $0.040^{+0.102}_{-0.115}$ & -- & -- & $68.75\pm2.45$ & -- \\
 & \hiig-2019 & $0.0224\pm0.0108$ & $0.1245^{+0.0413}_{-0.0380}$ & $0.285\pm0.077$ & $-0.052^{+0.289}_{-0.530}$ & -- & -- & $71.95\pm2.04$ & -- \\
 & \hiig-2021 & $0.0225\pm0.0108$ & $0.1035^{+0.0328}_{-0.0268}$ & $0.243^{+0.060}_{-0.055}$ & $-0.100^{+0.216}_{-0.484}$ & -- & -- & $72.15^{+2.05}_{-2.04}$ & -- \\
 & QSO & -- & $0.1797^{+0.0610}_{-0.1489}$ & $0.385^{+0.074}_{-0.191}$ & $0.043^{+0.265}_{-0.464}$ & -- & -- & $71.33^{+26.04}_{-9.82}$ & $11.27^{+0.91}_{-3.92}$ \\
 & HzBSNPD\tnote{c} & $0.0248^{+0.0036}_{-0.0043}$ & $0.1157\pm0.0164$ & $0.296\pm0.023$ & $0.027\pm0.074$ & -- & -- & $68.82\pm2.02$ & -- \\
 & HzBSNPDQH\tnote{d} & $0.0256^{+0.0035}_{-0.0042}$ & $0.1188\pm0.0138$ & $0.295\pm0.021$ & $0.011\pm0.067$ & -- & -- & $69.90\pm1.18$ & $10.96\pm0.25$ \\
\\
Flat XCDM & $H(z)$ + BAO & $0.0297^{+0.0046}_{-0.0053}$ & $0.0934^{+0.0195}_{-0.0169}$ & $0.283^{+0.023}_{-0.021}$ & -- & $-0.751^{+0.152}_{-0.106}$ & -- & $65.85^{+2.38}_{-2.65}$ & -- \\
 & \hiig-2019 & $0.0224\pm0.0109$ & $0.1491^{+0.0596}_{-0.0390}$ & $0.327^{+0.108}_{-0.077}$ & -- & $-1.494^{+0.858}_{-0.411}$ & -- & $72.37^{+2.18}_{-2.20}$ & -- \\
 & \hiig-2021 & $0.0223\pm0.0108$ & $0.1297^{+0.0493}_{-0.0308}$ & $0.288^{+0.087}_{-0.058}$ & -- & $-1.527^{+0.786}_{-0.391}$ & -- & $72.66\pm2.19$ & -- \\
 & QSO & -- & $0.1785^{+0.0536}_{-0.1603}$ & $0.373^{+0.072}_{-0.187}$ & -- & $-1.709^{+0.696}_{-0.882}$ & -- & $>39.72$ & $11.90^{+1.52}_{-4.17}$ \\
 & HzBSNPD\tnote{c} & $0.0256^{+0.0031}_{-0.0035}$ & $0.1121^{+0.0107}_{-0.0108}$ & $0.293\pm0.016$ & -- & $-0.935\pm0.063$ & -- & $68.57\pm1.74$ & -- \\
 & HzBSNPDQH\tnote{d} & $0.0267^{+0.0029}_{-0.0033}$ & $0.1142\pm0.0103$ & $0.290\pm0.016$ & -- & $-0.950\pm0.062$ & -- & $69.69\pm1.20$ & $10.94\pm0.25$ \\
\\
Non-flat XCDM & $H(z)$ + BAO & $0.0288^{+0.0049}_{-0.0054}$ & $0.0997^{+0.0210}_{-0.0211}$ & $0.294\pm0.027$ & $-0.112^{+0.136}_{-0.137}$ & $-0.706^{+0.135}_{-0.084}$ & -- & $66.01\pm2.43$ & -- \\
 & \hiig-2019 & $0.0223\pm0.0107$ & $0.1288^{+0.0548}_{-0.0620}$ & $0.291^{+0.109}_{-0.110}$ & $0.089^{+0.484}_{-0.283}$ & $-1.409^{+0.810}_{-0.402}$ & -- & $71.96^{+2.06}_{-2.29}$ & -- \\
 & \hiig-2021 & $0.0224\pm0.0107$ & $0.1109^{+0.0472}_{-0.0514}$ & $0.255\pm0.090$ & $0.078^{+0.417}_{-0.322}$ & $-1.461^{+0.777}_{-0.361}$ & -- & $72.23^{+2.08}_{-2.28}$ & -- \\
 & QSO & -- & $0.1905^{+0.0461}_{-0.1731}$ & $0.403^{+0.093}_{-0.225}$ & $0.170^{+0.403}_{-0.256}$ & $-1.455^{+1.014}_{-0.776}$ & -- & $>38.43$ & $11.08^{+1.19}_{-4.05}$ \\
 & HzBSNPD\tnote{c} & $0.0245^{+0.0036}_{-0.0043}$ & $0.1193\pm0.0170$ & $0.302\pm0.024$ & $-0.069\pm0.119$ & $-0.907^{+0.099}_{-0.062}$ & -- & $68.95\pm1.96$ & -- \\
 & HzBSNPDQH\tnote{d} & $0.0255^{+0.0035}_{-0.0041}$ & $0.1189^{+0.0136}_{-0.0135}$ & $0.297\pm0.020$ & $-0.054\pm0.096$ & $-0.926^{+0.091}_{-0.062}$ & -- & $69.73^{+1.19}_{-1.20}$ & $10.89\pm0.26$ \\
\\
Flat $\phi$CDM & $H(z)$ + BAO & $0.0323^{+0.0060}_{-0.0034}$ & $0.0810^{+0.0188}_{-0.0185}$ & $0.267\pm0.025$ & -- & -- & $1.530^{+0.644}_{-0.904}$ & $65.09^{+2.23}_{-2.24}$ & -- \\
 & \hiig-2019 & $0.0214^{+0.0084}_{-0.0128}$ & $0.0561^{+0.0157}_{-0.0541}$ & $0.155^{+0.047}_{-0.097}$ & -- & -- & $<7.803$ & $70.97^{+1.91}_{-1.89}$ & -- \\
 & \hiig-2021 & $0.0213^{+0.0080}_{-0.0130}$ & $0.0468^{+0.0152}_{-0.0431}$ & $0.135^{+0.043}_{-0.077}$ & -- & -- & $2.454^{+0.587}_{-2.434}$ & $71.15^{+1.90}_{-1.89}$ & -- \\
 & QSO & $0.0221^{+0.0179}_{-0.0171}$ & $<0.3644$ & $0.305^{+0.081}_{-0.212}$ & -- & -- & $4.526^{+1.546}_{-4.464}$ & $70.79^{+24.27}_{-12.67}$ & $10.32^{+1.08}_{-3.54}$ \\
 & HzBSNPD\tnote{c} & $0.0273^{+0.0032}_{-0.0039}$ & $0.1051^{+0.0122}_{-0.0103}$ & $0.284\pm0.017$ & -- & -- & $0.351^{+0.132}_{-0.284}$ & $68.33\pm1.81$ & -- \\
 & HzBSNPDQH\tnote{d} & $0.0284^{+0.0027}_{-0.0035}$ & $0.1078^{+0.0112}_{-0.0090}$ & $0.282\pm0.016$ & -- & -- & $0.288^{+0.098}_{-0.252}$ & $69.54\pm1.17$ & $10.92\pm0.25$ \\
\\
Non-flat $\phi$CDM & $H(z)$ + BAO & $0.0319^{+0.0061}_{-0.0037}$ & $0.0849^{+0.0178}_{-0.0217}$ & $0.271^{+0.025}_{-0.028}$ & $-0.074^{+0.104}_{-0.111}$ & -- & $1.646^{+0.680}_{-0.840}$ & $65.48\pm2.26$ & -- \\
 & \hiig-2019 & $0.0216^{+0.0089}_{-0.0121}$ & $0.0560^{+0.0198}_{-0.0476}$ & $0.157^{+0.047}_{-0.091}$ & $0.314^{+0.310}_{-0.182}$ & -- & $3.983^{+1.228}_{-3.932}$ & $70.53\pm1.79$ & -- \\
 & \hiig-2021 & $0.0210^{+0.0076}_{-0.0129}$ & $0.0407^{+0.0116}_{-0.0381}$ & $0.125^{+0.037}_{-0.069}$ & $0.302^{+0.278}_{-0.221}$ & -- & $<8.046$ & $70.49\pm1.81$ & -- \\
 & QSO & $0.0222\pm0.0109$ & $0.1444^{+0.0391}_{-0.1352}$ & $0.330^{+0.108}_{-0.183}$ & $0.207^{+0.244}_{-0.241}$ & -- & $4.773^{+2.487}_{-3.699}$ & $70.15^{+20.58}_{-15.20}$ & $10.26^{+1.24}_{-3.42}$ \\
 & HzBSNPD\tnote{c} & $0.0259^{+0.0037}_{-0.0043}$ & $0.1136\pm0.0156$ & $0.294\pm0.022$ & $-0.088^{+0.105}_{-0.075}$ & -- & $0.495^{+0.208}_{-0.344}$ & $68.84\pm1.88$ & -- \\
 & HzBSNPDQH\tnote{d} & $0.0267^{+0.0035}_{-0.0041}$ & $0.1141\pm0.0133$ & $0.290\pm0.020$ & $-0.072^{+0.074}_{-0.073}$ & -- & $0.405^{+0.165}_{-0.304}$ & $69.62\pm1.17$ & $10.87\pm0.26$ \\
\bottomrule
\end{tabular}%}
\begin{tablenotes}[flushleft]
\item [a] \hunit.
\item [b] pc.
\item [c] $H(z)$ + BAO + SN-Pantheon + SN-DES.
\item [d] $H(z)$ + BAO + SN-Pantheon + SN-DES + QSO + \hiig-2021.
\end{tablenotes}
\end{threeparttable}%
}
\end{sidewaystable*}

\subsection{Joint analyses results}
\label{subsec:Results_joint}

Since the constraints derived from $H(z)$, BAO, SN-Pantheon, SN-DES, QSO, \hiig-2019, and \hiig-2021 data are not mutually inconsistent, we jointly analyze combinations of these data and summarize these results in this subsection.

The $H(z)$ + BAO and HzBSNPD results are different from, but consistent with, what we obtained in \cite{Caoetal_2021b}. The differences arise from the different codes that we used to analyze the data; in \cite{Caoetal_2021b} we used \textsc{emcee}, whereas here we used \textsc{class} and \textsc{MontePython}. It is worth recalling here that, as mentioned above, \textsc{class} constrains \obhs\ in the range $0.00499 \leq \obh \leq 0.03993$. Therefore the parameter constraints differ more when the model and data prefer higher values of \obhs; this is especially true of the \pcdm\ model when it is fitted to the $H(z)$ + BAO data combination. As a result, the present constraints on \om\ and $\alpha$ in \pcdm\ with $H(z)$ + BAO data are higher and lower than the ones given in \cite{Caoetal_2021b}. The HzBSNPD results are, however, consistent.

The fit to the HzBSNPDQH data combination produces, for all models, the most interesting results. By adding QSO and \hiig-2021 data to HzBSNPD combination, the constraints are slightly tightened. Although the posterior means of \obhs\ and \ochs\ are relatively higher, those of \om\ are lower than the constraints from HzBSNPD. The \om\ constraints range from a low of $0.282\pm0.016$ (flat \pcdm) to a high of $0.298\pm0.013$ (flat \lcdm), a difference of only 0.78$\sigma$.

The constraints on $H_0$ are between $H_0=69.54\pm1.17$ \hunit\ (flat \pcdm) and $H_0=69.95\pm1.18$ \hunit\ (flat \lcdm) --- a difference of only 0.25$\sigma$ --- which are $0.64\sigma$ (flat \lcdm) and $0.51\sigma$ (flat \pcdm) higher than the median statistics estimate of $H_0=68 \pm 2.8$ \hunit\ \citep{chenratmed}, and $1.85\sigma$ (flat \lcdm) and $2.09\sigma$ (flat \pcdm) lower than the local Hubble constant measurement of $H_0 = 73.2 \pm 1.3$ \hunit\ \citep{Riess_2021}.\footnote{Other local expansion rate $H_0$ measurements result in slightly lower central values with slightly larger error bars \citep{rigault_etal_2015,zhangetal2017,Dhawan,FernandezArenas,Breuvaletal_2020, Efstathiou_2020, Khetan_et_al_2021,rameez_sarkar_2021, Freedman2021}. Our $H_0$ estimates are consistent with earlier median statistics determinations \citep{gott_etal_2001, Calabreseetal2012} as well as with other recent $H_0$ measurements \citep{chen_etal_2017,DES_2018,Gomez-ValentAmendola2018, planck2018b,dominguez_etal_2019,Cuceu_2019,zeng_yan_2019,schoneberg_etal_2019, Blum_et_al_2020, Lyu_et_al_2020, Philcox_et_al_2020, Birrer_et_al_2020, Denzel_et_al_2020,Pogosianetal_2020,Boruahetal_2020,Kimetal_2020,Harvey_2020, Zhang_Huang_2021,lin_ishak_2021}.}

For non-flat \lcdm, non-flat XCDM, and non-flat \pcdm, we find $\Omega_{\rm k_0}=0.011\pm0.067$, $\Omega_{\rm k_0}=-0.054\pm0.096$, and $\Omega_{\rm k_0}=-0.072^{+0.074}_{-0.073}$, respectively. The non-flat XCDM and \pcdm\ models favor closed geometry, while the non-flat \lcdm\ model favors open geometry. Note, however, that these results are all consistent with spatially flat hypersurfaces to within 1$\sigma$.

Our results show a slight preference for dark energy dynamics. For flat (non-flat) XCDM, $w_{\rm X}=-0.950\pm0.062$ ($w_{\rm X}=-0.926^{+0.091}_{-0.062}$), with central values being 0.81$\sigma$ (1.19$\sigma$) away from $w_{\rm X}=-1$; and for flat (non-flat) \pcdm, $\alpha=0.288^{+0.098}_{-0.252}$ ($\alpha=0.405^{+0.165}_{-0.304}$), with central values being 1.14$\sigma$ (1.33$\sigma$) away from $\alpha=0$.

The constraints on the nuisance parameter $l_{\rm m}$ are between $l_{\rm m}=10.87\pm0.26$ pc (non-flat \pcdm) and $l_{\rm m}=10.96\pm0.26$ pc (flat \lcdm), which differ by 0.24$\sigma$ and so are effectively model-independent, and consistent with $l_{\rm m}=11.03\pm0.25$ pc \citep{Cao_et_al2017b}.

\subsection{Model comparison}
\label{subsec:comparison}

The values of the reduced $\chi^2$ ($\chi^2/\nu$), $\Delta \chi^2$, $\Delta AIC$, and $\Delta BIC$ are reported in Table \ref{tab:BFP}, where $\Delta \chi^2$, $\Delta AIC$, and $\Delta BIC$, are the differences between the values of the $\chi^2$, $AIC$, and $BIC$ for a given model and the ones for flat \lcdm. Here a negative (positive) value of $\Delta \chi^2$, $\Delta AIC$, or $\Delta BIC$ means that the given statistic favors (disfavors) the model under consideration relative to flat \lcdm. We find that, except for a few of the $H(z)$ + BAO and \hiig-2019 cases, the flat \lcdm\ model is the most favored model among all six models we study. The $AIC$ does not show strong evidence against any of the models.\footnote{There is weak evidence for the reference model when $\Delta AIC(BIC) \in [0,2]$, positive evidence when $\Delta AIC(BIC) \in (2,6]$, strong evidence when $\Delta AIC(BIC) \in (6,10]$, and very strong evidence when $\Delta AIC(BIC) > 10$ \citep{Kass_Raftery}.} However, we find that some data combinations show strong evidence against the models we study, when these models are analyzed using the BIC, as follows. First, the HzBSNPD combination strongly disfavors non-flat \lcdm\ and very strongly disfavors non-flat XCDM and non-flat \pcdm. Second, the HzBSNPDQH combination strongly disfavors non-flat \lcdm, flat XCDM, and flat \pcdm, and very strongly disfavors non-flat XCDM and non-flat \pcdm. Furthermore, strong evidence against non-flat XCDM as well as non-flat \pcdm\ are provided by the \hiig-2021 and QSO data.

\section{Conclusion}
\label{sec:conclusion}

We find that the new \hiig-2021 data provide more restrictive cosmological parameter constraints and also prefer lower values of \om\!, \wx\!, and \ok\ than those favored by the \hiig-2019 data. 
 
We find that the QSO characteristic linear size $l_{\rm m}$ is relatively model-independent, so QSOs can be treated as approximate standard rulers but the uncertainty in $l_{\rm m}$ must be accounted for in the analysis.

We also jointly analyzed a total of 1411 measurements, consisting of 31 $H(z)$, 11 BAO, 1048 SN-Pantheon, 20 SN-DES, 120 QSO, and 181 \hiig-2021 data points to constrain cosmological and nuisance parameters in six flat and non-flat cosmological models. We can describe the relatively model-independent summary features of the constraints obtained from this $H(z)$ + BAO + SN-Pantheon + SN-DES + QSO + \hiig-2021 (HzBSNPDQH) data combination as follows.\footnote{The following summary values are obtained with the same method used in \cite{Caoetal_2021b}, where we take the summary central value to be the mean of the two of six central-most values. As for the uncertainty, we call the difference between the two central-most values twice the systematic uncertainty and the average of the two central-most error bars the statistical uncertainty, and compute the summary error bar as the quadrature sum of the two uncertainties.} First, the constraint on $l_{\rm m}$ is $l_{\rm m}=10.93\pm0.25$ pc, which is consistent with the $l_{\rm m}=11.03\pm0.25$ pc of  \cite{Cao_et_al2017b}. Second, the constraint on \om\ is $\Omega_{\rm m_0}=0.293 \pm 0.021$, which is in good agreement with many other recent measurements (e.g. $0.315\pm0.007$ from TT,TE,EE+lowE+lensing CMB anisotropy data in the flat \lcdm\ model of  \citealp{planck2018b}). Third, the determination of $H_0$ is $H_0=69.7\pm1.2$ \hunit, which is in better agreement with the estimate of \cite{chenratmed} than with the measurements of \cite{planck2018b} and \cite{Riess_2021}. There is some room for dark energy dynamics or a little spatial curvature energy density, but overall the flat \lcdm\ model is the best candidate model.

\section*{Acknowledgements}

We thank Javier de Cruz P\'{e}rez, Ana Luisa Gonz\'{a}lez-Mor\'{a}n, Narayan Khadka, and Chan-Gyung Park for useful discussions. This work was partially funded by Department of Energy grant DE-SC0011840. The computing for this project was performed on the Beocat Research Cluster at Kansas State University, which is funded in part by NSF grants CNS-1006860, EPS-1006860, EPS-0919443, ACI-1440548, CHE-1726332, and NIH P20GM113109.

\section*{Data availability}

The \hiig\ data used in this article were provided to us by the authors of \cite{G-M_2019, GM2021}. These data will be shared on request to the corresponding author with the permission of the authors of \cite{G-M_2019, GM2021}.

%%%%%%%%%%%%%%%%%%%%%%%%%%%%%%%%%%%%%%%%%%%%%%%%%%

%%%%%%%%%%%%%%%%%%%% REFERENCES %%%%%%%%%%%%%%%%%%

% The best way to enter references is to use BibTeX:

\bibliographystyle{mnras}
\bibliography{mybibfile} % if your bibtex file is called example.bib

% Alternatively you could enter them by hand, like this:
% This method is tedious and prone to error if you have lots of references
%\begin{thebibliography}{99}
%\bibitem[\protect\citeauthoryear{Author}{2012}]{Author2012}
%Author A.~N., 2013, Journal of Improbable Astronomy, 1, 1
%\bibitem[\protect\citeauthoryear{Others}{2013}]{Others2013}
%Others S., 2012, Journal of Interesting Stuff, 17, 198
%\end{thebibliography}

%%%%%%%%%%%%%%%%%%%%%%%%%%%%%%%%%%%%%%%%%%%%%%%%%%

%%%%%%%%%%%%%%%%% APPENDICES %%%%%%%%%%%%%%%%%%%%%

%\appendix

% If you want to present additional material which would interrupt the flow of the main paper,
% it can be placed in an Appendix which appears after the list of references.

%%%%%%%%%%%%%%%%%%%%%%%%%%%%%%%%%%%%%%%%%%%%%%%%%%

% Don't change these lines
\bsp	% typesetting comment
\label{lastpage}
\end{document}